# Hacia una moderna república de las ideas vía un nuevo ecosistema de comunicación científica

Enrique López González
Orcid: 0000-0003-1477-5416
Universidad de León (Dpto. Dirección y Economía de la Empresa)
Campus de Vegazana, s/n. 24071 - León (España)
enrique.lopez@unileon.es / enrique@lopezgonzalez.org

## Contenido

> *"Yo tengo mucha más lástima de un hombre que quiere saber y no puede, que de un hambriento. Porque un hambriento puede calmar su hambre fácilmente con un pedazo de pan o con unas frutas, pero un hombre que tiene ansia de saber y no tiene medios, sufre una terrible agonía porque son libros, libros, muchos libros los que necesita y ¿dónde están esos libros? ¡Libros! ¡Libros! Hace aquí una palabra mágica que equivale a decir: 'amor, amor', y que debían los pueblos pedir como piden pan o como anhelan la lluvia para sus sementeras".*
>
> Federico García Lorca (Discurso inauguración de la biblioteca pública de Fuente Vaqueros, 1931) [1]

> *"Todos los imperios del futuro serán imperios del conocimiento, y solamente los pueblos que entiendan cómo generar conocimiento y cómo protegerlo, cómo buscar jóvenes que tengan capacidad para hacerlo y asegurarse de que se queden en el país, serán países exitosos. Los otros, por más que tengan recursos materiales, materias primas diversas, litorales extensos, historias fantásticas, etc., probablemente no se queden ni con las mismas banderas, ni con las mismas fronteras, ni mucho menos con un éxito económico"*
>
> Albert Einstein.

# Introducción: Advenimiento de una nueva era de la comunicación científica.

El ecosistema académico contemporáneo, heredero de la "República de las Letras" [2] de la Ilustración, se encuentra en un estado de crisis profunda e insostenible. Aquel orden, basado en la libre circulación de la correspondencia y la búsqueda desinteresada del conocimiento, ha sido suplantado por un sistema que se tambalea bajo el peso de sus propias contradicciones. La presentación de la demanda antimonopolio *Uddin contra Elsevier* no es un mero litigio; es un punto de inflexión histórico, un síntoma que revela el fracaso de un pacto social obsoleto.

Durante décadas, hemos sido testigos de cómo la infraestructura de la comunicación científica ha sido capturada por un oligopolio extractivo que impone márgenes de beneficio anómalos sobre un bien público. Hemos observado cómo los incentivos del sistema han corrompido su propósito, generando una alarmante "crisis de replicación" que socava la fiabilidad misma del conocimiento que producimos. Y hemos permanecido paralizados, atrapados en un dilema de acción colectiva donde las instituciones, actuando racionalmente en su propio interés a corto plazo, perpetúan un sistema colectivamente irracional.

---

[1] https://www.cervantesvirtual.com/obra-visor/alocucion-al-pueblo-de-fuente-vaqueros-discurso-leido-por-la-inaguracion-de-la-biblioteca-publica-de-fuente-vaqueros-septiembre-1931-998622/html/a5692ac7-3664-4749-84da-9837f987e46d_2.html

[2] https://es.wikipedia.org/wiki/Rep%C3%BAblica_de_las_letras

Las reformas incrementales han fracasado. El momento no pide un ajuste; exige una refundación.

Este trabajo se embarca en una tarea de rediseño fundamental. Su objetivo no es proponer otra solución parcial, sino articular un marco original, innovador y coherente para un nuevo orden mundial en la comunicación científica: la fundación de una moderna República de las Ideas.

Para llevar a cabo tarea monumental, este análisis sintetiza cuatro corrientes de pensamiento distintas, pero profundamente complementarias. Del Ordoliberalismo tomamos el rigor de diseñar una "constitución económica" (*Ordnungspolitik*) que impida la concentración de poder y fomente la competencia leal. De la Economía Humanista, extraemos el *telos* o propósito normativo: un sistema diseñado no para la extracción de rentas, sino para el florecimiento humano y la prosperidad compartida. Del Humanismo Digital, derivamos el *ethos* tecnológico, asegurando que la infraestructura sirva a la dignidad y autonomía humanas, y no al revés. Finalmente, de la Ciencia Descentralizada (DeSci), tomamos el conjunto de herramientas arquitectónicas —contratos inteligentes, DAOs, tokens— capaces de construir este nuevo orden desde los cimientos.

El arco narrativo de este informe es deliberado: La Parte I ofrece una anatomía de la decadencia, utilizando la demanda *Uddin* como bisturí para diseccionar las patologías económicas, institucionales y epistémicas del sistema actual. La Parte II articula la constitución filosófica de la nueva república. La Parte III detalla el proyecto arquitectónico y la infraestructura tecnológica de este nuevo orden. La Parte IV somete este diseño a rigurosas pruebas de estrés para forjar su resiliencia y antifragilidad. La Parte V traza una hoja de ruta estratégica para la transición, un camino plausible para navegar desde el *statu quo* hasta la plena soberanía comunitaria. Finalmente, la Parte VI cierra el círculo, volviendo a la visión fundacional. No es un resumen, sino un manifiesto constituyente, una llamada a la acción para construir la Nueva República de las Ideas que la búsqueda de la verdad merece en el siglo XXI.

# I. La Anatomía de la Decadencia: El Caso Incontrovertible para la Reconstrucción Sistémica

El ecosistema académico contemporáneo no sufre de problemas aislados, sino de una falla sistémica integral. Las patologías que lo aquejan —económicas, institucionales y epistémicas— no son disfunciones independientes, sino facetas interconectadas de un único diseño institucional fundamentalmente quebrado. Este diagnóstico revela que la reforma incremental es fútil y la reconstrucción fundamental, una necesidad ineludible.

La demanda colectiva presentada en septiembre de 2024 por la Dra. Lucina Uddin, profesora de neurociencia de la UCLA, contra seis de las mayores editoriales académicas del mundo —*Elsevier, Springer Nature, Taylor and Francis, Sage, Wiley y Wolters Kluwer*— no es simplemente una disputa legal sobre prácticas comerciales [3]. Representa la formalización jurídica de décadas de frustración acumulada dentro de la comunidad científica. Transforma lo que durante mucho tiempo se ha descrito como un "sistema roto" o éticamente cuestionable en una acusación formal de una "conspiración ilegal" bajo la ley antimonopolio (Ley Sherman) de los Estados Unidos [4], [5].

La presentación de la demanda antimonopolio *Uddin contra Elsevier* es el punto de inflexión que ilumina las fracturas estructurales de un sistema que ha priorizado la extracción de rentas sobre la creación de conocimiento, la señalización de prestigio sobre la búsqueda de la verdad y los intereses privados sobre el bien público.

---

[3] https://www.lieffcabraser.com/antitrust/academic-journals/

[4] https://www.findlaw.com/legalblogs/courtside/academic-publishers-seek-to-exit-war-over-words-with-scientific-professors/

[5] https://academic.oup.com/healthaffairsscholar/article/3/2/qxaf018/8002321

# I.1 El Oligopolio Editorial como Patología Económica

La estructura del mercado de la publicación académica [6] representa una anomalía económica extrema. Lejos de ser un campo de competencia vibrante, se ha consolidado en un oligopolio estable y altamente rentable, dominado por un número reducido de corporaciones. Cinco grandes editoriales —*Elsevier, John Wiley & Sons, Taylor & Francis, Springer Nature y SAGE*— acaparan más del 50% de los ingresos de un mercado global valorado en más de 19 mil millones de dólares anuales [7]. Esta concentración de poder de mercado permite a estas empresas ejercer una capacidad de fijación de precios extraordinaria sobre un bien cuya demanda es inelástica: el acceso a la investigación de vanguardia, financiada en gran medida con fondos públicos, es indispensable para el funcionamiento de las universidades y el avance de la ciencia.

El resultado es una industria con márgenes de beneficio que son, a todas luces, anómalos: las grandes editoriales académicas reportan consistentemente márgenes de beneficio operativo en el rango del 30% al 40% [8]. Así, en 2018, por ejemplo, la división de ciencia, tecnología y medicina de *Elsevier* registró un margen de beneficio de más del 40% [9]. Esta rentabilidad extraordinaria no es un signo de eficiencia o de una creación de valor superior, sino una clara indicación de un poder de mercado inmenso.

La lucratividad del modelo se deriva directamente de la externalización de sus costes más significativos —la creación y validación del contenido— a la comunidad académica y, en última instancia, a los contribuyentes. En esencia, el oligopolio editorial ha transformado un bien público —el conocimiento— en un activo privado generador de rentas, funcionando en la práctica como un "impuesto privado sobre el proceso de creación de conocimiento". De hecho, los beneficios exorbitantes representan una transferencia masiva de riqueza desde el sector público y sin ánimo de lucro hacia los accionistas privados.

Este mecanismo de extracción ha creado lo que el *Deutsche Bank* [10] describió como un "bizarro sistema de triple pago", que opera a través de una triple apropiación de recursos, en gran parte financiados con fondos públicos [11], a saber:

- Primero, los contribuyentes, a través de subvenciones gubernamentales a las entidades públicas, financian la investigación científica.
- Segundo, los académicos, cuyo trabajo es sostenido por estas mismas instituciones, realizan la investigación, escriben los manuscritos y, crucialmente, llevan a cabo la revisión por pares —el control de calidad esencial del sistema— sin compensación directa de las editoriales, externalizando así sus costes operativos principales.
- Tercero, las mismas instituciones financiadas con fondos públicos (bibliotecas universitarias, centros de investigación) deben pagar de nuevo para acceder a los resultados de esa misma investigación, la que ellas mismas financiaron, a través de suscripciones a revistas con precios exorbitantes y crecientes barreras de pago.

El resultado, según la demanda, no es solo un enriquecimiento injusto, sino un freno deliberado al avance de la ciencia. Además, este sistema se perpetúa por la dinámica de "publicar o perecer" intrínseca a la carrera académica [12]: para los investigadores, la publicación en revistas de alto prestigio, controladas predominantemente por las editoriales demandadas, no es una opción, sino una necesidad para la obtención de empleo, la promoción, la titularidad y la financiación de futuras investigaciones. Esta dependencia crea un mercado de compradores cautivos para las suscripciones de las bibliotecas y un suministro de mano de obra cautiva para la escritura y la revisión. La demanda argumenta que las editoriales explotan esta dependencia estructural para imponer términos que serían insostenibles en un mercado competitivo, transformando el avance de la carrera académica en el motor de su rentabilidad [13].

---

[6] https://southernlibrarianship.icaap.org/content/v09n03/mcguigan_g01.html

[7] https://wordsrated.com/academic-publishers-statistics/

[8] https://royalsociety.org/blog/2015/05/fssc-are-publisher-profits-justifiable/

[9] https://pmc.ncbi.nlm.nih.gov/articles/PMC6740196/

[10] https://www.lieffcabraser.com/pdf/AcademicPublicationsComplaintFinal.pdf

[11] https://www.bmj.com/content/386/bmj.q2037

[12] https://journals.library.columbia.edu/index.php/lawandarts/announcement/view/742

## I.2 La Demanda Uddin vs. Elsevier como Herramienta de Diagnóstico

La demanda colectiva antimonopolio presentada en nombre de la Dra. Lucina Uddin expone los mecanismos de control que sostienen esta estructura de mercado disfuncional [14]. Las acusaciones centrales no se limitan a la fijación de precios, sino que atacan la arquitectura operativa de la industria [15].

Un esquema que se sustenta en tres "acuerdos" colusorios interconectados que, en conjunto, neutralizan la competencia en los mercados de trabajo intelectual, manuscritos y diseminación del conocimiento, a saber:
- La Regla de la Revisión por Pares No Remunerada *(Unpaid Peer Review Rule)*
- La Regla de la Sumisión Única *(Single Submission Rule)*
- La "Ley Mordaza" y la Apropiación de la Propiedad Intelectual *(Gag Rule)*

### I.2.1 Fijación de Precios del Trabajo Intelectual Cero: La Regla de la Revisión por Pares No Remunerada (Unpaid Peer Review Rule)

La primera y quizás más flagrante alegación es que los editores demandados han conspirado para fijar el precio de los servicios de revisión por pares en cero. La revisión por pares es el cimiento de la validación científica, un servicio especializado y de alto valor sin el cual las publicaciones académicas serían "esencialmente inútiles". En un mercado competitivo, se esperaría que las editoriales compitieran por los servicios de revisores expertos ofreciendo alguna forma de compensación. Sin embargo, la demanda alega que existe un acuerdo horizontal entre los demandados para no pagar por este trabajo, lo que constituye una forma clásica de fijación de precios de un insumo laboral, una violación *per se* de la ley antimonopolio [16].

Este acuerdo, según la demanda, no es pasivo, sino que se impone mediante un mecanismo de coerción. Se alega que las editoriales vinculan implícita y explícitamente la voluntad de un académico de realizar revisiones no remuneradas con su capacidad para que sus propios manuscritos sean publicados en las prestigiosas revistas de los demandados.

En el entorno de "publicar o perecer", esta vinculación convierte la provisión de trabajo gratuito en una condición no escrita para la supervivencia profesional, "manteniendo las carreras de los académicos como rehenes". Este sistema de trueque forzado de trabajo por oportunidad de publicación suprime cualquier mercado potencial para los servicios de revisión y transfiere miles de millones de dólares en valor laboral de la comunidad académica a los beneficios de las editoriales.[2]

---

[13] https://docs.lib.purdue.edu/clcweb/vol16/iss1/12/

[14] https://storage.courtlistener.com/recap/gov.uscourts.nyed.520652/gov.uscourts.nyed.520652.63.0.pdf

[15] https://www.promarket.org/2024/04/24/high-prices-and-market-power-of-academic-publishing-reduce-article-citations/

[16] https://pmc.ncbi.nlm.nih.gov/articles/PMC11823101/

## I.2.2 Supresión de la Competencia: La Regla de la Sumisión Única (Single Submission Rule)

La segunda columna de la conspiración alegada es la "Regla de Sumisión Única", un acuerdo por el cual las editoriales exigen a los autores que envíen sus manuscritos a una sola revista a la vez. Se otorga así un monopolio temporal a la primera revista que recibe un manuscrito, eliminando los incentivos para competir en velocidad o calidad de la revisión y ralentizando artificialmente el progreso científico. Si el manuscrito es rechazado, a menudo después de un largo proceso de revisión, el autor puede entonces enviarlo a otra revista. La demanda argumenta que esta práctica es un acuerdo anticompetitivo para no competir entre sí por el "activo" más valioso de la industria: los manuscritos científicos.

En un mercado funcional, un autor podría enviar su trabajo a varias revistas simultáneamente, permitiendo que estas compitan en base a la velocidad y calidad de la revisión, los términos de publicación y el prestigio ofrecido. La Regla de Sumisión Única elimina esta competencia. Reduce drásticamente los incentivos para que las revistas procesen los manuscritos de manera eficiente, ya que no enfrentan el riesgo de que un competidor más rápido publique el trabajo primero. Además, una vez que una revista ha invertido tiempo en revisar un manuscrito y ofrece su publicación, el autor se encuentra en una posición de negociación extremadamente débil. Habiendo esperado meses o incluso más de un año, la alternativa de retirar el manuscrito y comenzar el proceso de nuevo en otra revista es a menudo profesionalmente inviable. Esto permite a la editorial "dictar los términos de la publicación", incluyendo la cesión de los derechos de autor.

La demanda enmarca esta regla como un acuerdo horizontal para limitar la producción y la innovación, lo que restringe el comercio de forma irrazonable.

## I.2.3 Restricción del Conocimiento: La "Ley Mordaza" y la Apropiación de la Propiedad Intelectual (Gag Rule)

El tercer componente del esquema es la "Ley Mordaza" (*Gag Rule*), que prohíbe a los académicos compartir o discutir libremente los avances científicos descritos en sus manuscritos desde el momento de la sumisión hasta la publicación, un proceso que puede durar meses o incluso años.

La demanda argumenta que, desde el instante en que se envía un manuscrito, las editoriales se comportan como si los avances científicos contenidos en él fueran de su propiedad, para ser compartidos solo con su permiso.

Esta restricción se ve agravada por la práctica de exigir a los autores que cedan la totalidad de sus derechos de propiedad intelectual a la editorial como condición para la publicación, a menudo a cambio de nada. El conocimiento, frecuentemente generado con fondos públicos, se convierte así en propiedad privada de la editorial. Esta puede entonces cobrar "*el máximo que el mercado pueda soportar*" por el acceso a dicho conocimiento, vendiéndolo de vuelta a la misma comunidad que lo produjo.

La demanda alega que este acuerdo para restringir la libre circulación del conocimiento científico y para apropiarse de la propiedad intelectual sin compensación es una restricción irrazonable del comercio que va en contra del propósito fundamental de la ciencia: la rápida y amplia diseminación del saber.

## I.2.4 Consecuencias Sistémicas: La Crisis de la Revisión por Pares y el Freno al Progreso Científico

Las prácticas alegadas en la demanda no solo generan beneficios desmesurados para las editoriales, sino que también producen fallos de mercado perversos que perjudican la empresa científica en su totalidad. La negativa a compensar a los revisores, combinada con la creciente carga de trabajo de los académicos, ha precipitado una "crisis de la revisión por pares". Cada vez es más difícil encontrar expertos dispuestos a dedicar su valioso tiempo a realizar revisiones rigurosas de forma gratuita. Como resultado, los manuscritos pueden permanecer en espera de revisión durante meses o incluso años, creando un cuello de botella que retrasa la comunicación de hallazgos importantes.

Este retraso sistémico tiene consecuencias profundas. La demanda argumenta que el esquema "*ha frenado la ciencia, retrasando los avances en todos los campos de la investigación*". La implicación es directa y grave: se tardará más en encontrar tratamientos eficaces para el cáncer, más tiempo en desarrollar tecnologías para combatir el cambio climático y más tiempo en lograr cualquier avance que dependa de la rápida difusión y validación del conocimiento científico.

Al traducir este coste de oportunidad en un daño legalmente reconocible, la demanda *Uddin* intenta cuantificar el precio que la sociedad paga por un modelo de publicación que, según alega, prioriza el beneficio del oligopolio sobre el progreso del conocimiento humano.

El modelo de negocio de las editoriales académicas puede entenderse no solo como una serie de acuerdos separados, sino como una forma sofisticada de "empaquetamiento" (*bundling*) anticompetitivo [17]. En la teoría antimonopolio, el empaquetamiento ilegal ocurre cuando una empresa con poder de monopolio sobre un producto (el producto "vinculante" o *tying product*) obliga a los clientes a comprar también un segundo producto (el producto "vinculado" o *tied product*) para poder obtener el primero [18]. Así, en el caso de la publicación académica, el producto vinculante, donde las grandes editoriales ejercen un poder de mercado casi absoluto, es un bien intangible pero indispensable: el prestigio o reputación [19].

La publicación en una revista de élite de *Elsevier* o *Springer Nature* es la credencial fundamental para el avance profesional en la academia. Las editoriales aprovechan este "monopolio del prestigio" para empaquetar un conjunto de términos y servicios vinculados que los académicos se ven obligados a aceptar. Para acceder al producto "prestigio", un autor debe aceptar el paquete completo: ceder su manuscrito en exclusiva (Regla de Sumisión Única), proporcionar trabajo de revisión gratuito (Regla de Revisión por Pares No Remunerada) y, finalmente, transferir su propiedad intelectual (Ley Mordaza y cesión de derechos).

Este análisis, que se inspira en casos antimonopolio como *LePage's Inc. v. 3M* [20], [21], sugiere que las editoriales no están simplemente vendiendo acceso a revistas; están vendiendo acceso a la progresión profesional y utilizando ese poder de mercado para imponer condiciones anticompetitivas que excluyen modelos de publicación alternativos, más eficientes o justos para los autores.

La demanda puede o no tener éxito en los tribunales, pero al menos tiene el valor de haber catalizado una conversación ineludible sobre la necesidad de una alternativa.

---

[17] https://www.wsgr.com/a/web/178/jacobson-07.pdf

[18] https://www.justice.gov/archives/atr/antitrust-analysis-bundled-loyalty-discounts

[19] https://docs.lib.purdue.edu/clcweb/vol16/iss1/12/

[20] https://nyulawreview.org/issues/volume-79-number-4/lepages-v-3m-an-antitrust-analysis-of-loyalty-rebates/

[21] https://www.law.berkeley.edu/wp-content/uploads/2015/04/LePages-v-3M-2003-excerpt.pdf

# I.3 La Crítica "Peculiar por Diseño": Desenmascarando el Verdadero Foco del Fracaso

Aunque la demanda *Uddin* articula con fuerza las quejas de la comunidad académica, un análisis más profundo revela que su estructura es "*peculiar por diseño*" [22]. La demanda enmarca el conflicto como una disputa bilateral exclusiva entre académicos y editoriales. Sin embargo, esta formulación excluye convenientemente a los actores que tradicionalmente han financiado la mayor parte de los crecientes costes de publicación: las universidades y los organismos de financiación gubernamentales. Son estas instituciones, respaldadas por el dinero de los contribuyentes, las que soportan la carga de las suscripciones a revistas. Sin embargo, están ausentes en la ecuación legal analizada.

Los abogados de los demandantes han tomado prestados fragmentos del lenguaje familiar utilizado para describir el "Dilema de la Comunicación Académica", pero lo han reconducido hacia un objetivo específico y limitado, enfocado apropiadamente en una demanda colectiva antimonopolio: *"aumentar la eficiencia del capitalismo eliminando esos obstáculos"*. Se evidencia así que su objetivo no es una reforma sistémica, sino la eliminación de una distorsión del mercado para que los académicos (como clase) puedan, en teoría, ser compensados por su trabajo.

Esta simplificación estratégica, si bien es comprensible desde una perspectiva legal, oculta una verdad mucho más profunda sobre la naturaleza del problema. Al centrarse únicamente en la mala conducta de las editoriales, la demanda ignora la pregunta fundamental: ¿por qué y cómo estas editoriales llegaron a ostentar un poder de mercado tan inmenso?

La respuesta no se encuentra en la fortaleza de las editoriales, sino en la debilidad estructural de las instituciones académicas.

# I.4 La Causa Raíz: La Comunicación Académica como un Dilema de Acción Colectiva

La persistencia del oligopolio editorial, a pesar de su modelo de negocio evidentemente extractivo, no puede explicarse únicamente por el poder de mercado de las editoriales. ¿Acaso pudiera sustentarse en una parálisis sistémica dentro de la propia comunidad académica?

La paradoja central de la comunicación académica es que el auge de Internet, que ha reducido drásticamente el coste de distribución de la información a casi cero, ha coincidido con aumentos drásticos en los precios que las bibliotecas académicas pagan por el acceso a las revistas científicas. Esta aparente contradicción se resuelve a través de la lente de la "teoría de la acción colectiva" [23] del economista Mancur Olson, quien en su obra seminal *The Logic of Collective Action: Public Goods and the Theory of Groups (1965)* [24], [25] argumentó que, incluso si todos los individuos de un grupo grande son racionales y se beneficiarían de actuar juntos para lograr un objetivo o interés común, a menudo no lo harán voluntariamente debido al incentivo del "free-riding" o parasitismo.

Este dilema se manifiesta de forma aguda en el ecosistema académico: el sistema de comunicación académica es un ejemplo de manual de este dilema.

El bien público deseado es un ecosistema de conocimiento abierto y asequible, pero los participantes están atrapados en una red de incentivos perversos que impiden la cooperación, a saber:

El dilema de los investigadores no es baladí. La progresión profesional en la academia —la obtención de la titularidad, la consecución de subvenciones y el reconocimiento general— depende de la publicación en revistas de alto prestigio. Estas revistas, que actúan como el principal mecanismo de señalización de la reputación académica, están abrumadoramente controladas por el oligopolio editorial [26]. Para un investigador individual, boicotear estas revistas de élite equivale a un autosabotaje profesional. Aunque un boicot colectivo por parte de toda la comunidad científica podría quebrar el poder de las editoriales y beneficiar a todos a largo plazo, el coste a corto plazo para cualquier individuo que actúe solo es prohibitivamente alto. Esta es la esencia de la cultura de "publicar o perecer".

---

[22] https://www.charleston-hub.com/2025/07/uddin-v-elsevier-peculiar-by-design/

[23] https://digitalcommons.unl.edu/cgi/viewcontent.cgi?article=1041&context=scholcom

[24] http://commres.net/wiki/_media/olson.pdf

[25] https://sociologia1unpsjb.wordpress.com/wp-content/uploads/2008/03/olson-logica-accion-colectiva.pdf

El dilema de las instituciones es parejo. Las universidades y sus bibliotecas no pueden cancelar unilateralmente las suscripciones de "grandes paquetes" a las revistas esenciales de editoriales como Elsevier sin paralizar la capacidad de investigación de sus propios profesores y estudiantes. Se ven obligadas a pagar precios que aumentan anualmente a un ritmo muy superior al de la inflación, temiendo que, si cancelan, otras instituciones no lo harán, dejándolas en una desventaja competitiva significativa [27]. Esta dinámica ha alimentado la "crisis de las publicaciones seriadas", un fenómeno de décadas de duración en el que los costes de las revistas han consumido una parte cada vez mayor de los presupuestos de las bibliotecas, desplazando la compra de otros recursos vitales como las monografías.

Este no es un problema de las bibliotecas, sino una crisis para toda la "empresa" académica que se manifiesta a través de los presupuestos de las bibliotecas. Representa una masiva y sistémica mala asignación de capital, desviando fondos de la investigación primaria y la educación hacia los beneficios de los accionistas, socavando directamente la misión central de las universidades.

El bien común deseado es un sistema de comunicación científica [28] abierto, eficiente y de bajo coste, una infraestructura de conocimiento compartida gestionada por la propia comunidad académica. Sin embargo, la creación de este bien público se ve socavada por los incentivos individuales. Así, una universidad específica, actuando en su propio interés racional, carece del incentivo necesario para invertir en una colección compartida. Se enfrenta a la tentación de convertirse en un "free rider" (polizón), esperando que otras instituciones asuman los costes de construcción de la infraestructura mientras ella disfruta de los beneficios. Incluso si una institución no tiene la intención de ser un polizón, duda en invertir por temor a que su esfuerzo se desperdicie si otras muchas eligen no contribuir.

Este fracaso de la coordinación, esta incapacidad de las universidades para actuar colectivamente en su propio interés a largo plazo fue lo que creó un vacío de poder. Las editoriales comerciales no surgieron como depredadores externos, sino como una solución de mercado a este dilema de acción colectiva. Se convirtieron en los gestores externalizados del sistema de comunicación científica, asumiendo las funciones de coordinación, revisión y distribución que la comunidad académica no pudo organizar por sí misma. Con el tiempo, esta posición les permitió desarrollar un poder de mercado oligopólico, transformando un servicio en un mecanismo de extracción de rentas a gran escala.

En definitiva, el poder de las editoriales no es la causa de la crisis, sino un síntoma de la parálisis institucional subyacente. La demanda *Uddin*, al ignorar a las universidades, apunta al síntoma mientras deja intacta la enfermedad.

---

[26] https://en.wikipedia.org/wiki/Academic_publishing
[27] https://southernlibrarianship.icaap.org/content/v09n03/mcguigan_g01.html
[28] https://en.wikipedia.org/wiki/Academic_publishing

# I.5 Contexto Histórico: La Comercialización de las Sociedades Científicas

La situación denunciada no surgió de la noche a la mañana. Es el resultado de una transformación histórica. La publicación académica se originó en el seno de sociedades científicas sin ánimo de lucro, como la *Royal Society* en el siglo XVII, cuya misión era la difusión del conocimiento. Para estas sociedades, la publicación era un medio para compartir el conocimiento con la comunidad académica y generar prestigio institucional; la reutilización y la reimpresión ayudaban a lograr ese objetivo.

El punto de inflexión se produjo en las décadas de 1960 y 1970, cuando las editoriales comerciales comenzaron a adquirir selectivamente las revistas de "máxima calidad" que antes publicaban las sociedades académicas sin ánimo de lucro [29]. Al aumentar significativamente los precios de las suscripciones, estas editoriales comerciales descubrieron que perdían poco mercado debido a la demanda inelástica de estas prestigiosas publicaciones.

El mecanismo legal que consolidó este control fue la introducción de los acuerdos de transferencia de derechos de autor. A partir de finales del siglo XX, se convirtió en práctica habitual que los académicos cedieran gratuitamente los derechos de autor de sus investigaciones financiadas con fondos públicos a las editoriales como condición para su publicación [30].

Este cambio transformó la economía de la publicación académica. La economía del prestigio de la academia se convirtió en el mecanismo de aplicación del dilema de la acción colectiva. Las editoriales no se limitan a vender el acceso a los contenidos; controlan el principal mecanismo de señalización de la reputación académica. Este "bloqueo reputacional" es más poderoso que cualquier barrera tecnológica o económica, ya que obliga a actores racionales a participar en un sistema que es colectivamente irracional.

El poder de mercado de las editoriales no es, por tanto, puramente económico; está profundamente entrelazado con los sistemas de recompensa social y cultural de la propia ciencia. En consecuencia, cualquier reforma exitosa debe no solo proporcionar un modelo económico alternativo, sino también un sistema de prestigio alternativo y viable.

# I.6 La Consecuencia: La Crisis de Replicación y la Erosión de la Confianza Epistémica

El coste de este fracaso sistémico no es meramente económico; es profundamente epistémico. La defectuosa arquitectura institucional inflige un daño directo a la integridad de la propia ciencia: la crisis de replicación o reproducibilidad. Esta crisis no es un problema de mala ciencia por parte de individuos, sino el resultado inevitable de un sistema de incentivos que recompensa la novedad por encima del rigor y la publicación por encima de la verdad.

La estructura de mercado disfuncional ha creado un sistema de incentivos perversos que corrompe el proceso científico mismo. Por un lado, la cultura de "publicar o perecer", en la que la carrera de un académico depende de la publicación en un pequeño número de revistas de "alto impacto" controladas por el oligopolio editorial, fomenta directamente la "crisis de replicación" [31]. Por otro lado, las revistas prestigiosas, para justificar sus altos costes y su exclusividad, tienen un fuerte sesgo de publicación hacia resultados "innovadores", positivos, "sorprendentes" y estadísticamente significativos [32].

---

[29] https://open-access.network/en/information/open-access-primers/history-of-the-open-access-movement

[30] https://blogs.lse.ac.uk/impactofsocialsciences/2019/06/03/what-the-history-of-copyright-in-academic-publishing-tells-us-about-open-research/

[31] https://www.news-medical.net/life-sciences/What-is-the-Replication-Crisis.aspx

[32] https://replicationindex.com/2015/01/24/qrps/

De esta forma, se crea un fuerte desincentivo para publicar hallazgos nulos o estudios de replicación exitosa de un estudio anterior se consideran menos publicables, inclinando el registro científico público hacia los falsos positivos. O dicho a la inversa, a modo de palíndromo, esta presión crea un incentivo abrumador para que los investigadores adopten prácticas de investigación cuestionables. Entre las más comunes se encuentran el *p-hacking*, que consiste en manipular los datos o los análisis estadísticos hasta que un resultado no significativo se vuelve significativo, y el HARKing (*Hypothesizing After the Results are Known*), que implica formular una hipótesis después de haber visto los resultados para que parezca predictiva [33].

Las consecuencias son devastadoras para la fiabilidad del conocimiento científico. Las cifras son alarmantes. Una encuesta de *Nature* reveló que más del 70% de los investigadores habían intentado y no habían podido reproducir los resultados de otro científico [34]. Un estudio de 2015 en la revista Science reveló que solo el 36% de los experimentos en psicología podían replicarse [35] Un estudio publicado en *Nature Human Behavior* no logró replicar 13 de 21 artículos de ciencias sociales y del comportamiento publicados en *Science* y *Nature*, dos de las revistas más importantes del mundo [36]. En el campo de la investigación biomédica, un estudio de 2012 descubrió que solo el 11% de 53 estudios preclínicos sobre cáncer podían replicarse con éxito. El coste económico de este fracaso es catastrófico: se estima que la investigación preclínica irreproducible en Estados Unidos cuesta aproximadamente 28 mil millones de dólares al año, un despilfarro colosal de fondos públicos y filantrópicos [37].

Para colmo de males, el sistema de recompensas académicas está completamente invertido. Paradójicamente, la evidencia del estudio de 2021 llevado a cabo por Marta Serra-García y Uri Gneezy [38] sugiere que los artículos con resultados no replicables tienden a ser citados con más frecuencia que los que sí lo son. A la inversa, que los artículos con resultados reproducibles tienden a ser menos citados que los artículos con hallazgos que no se pueden replicar. Esto se debe probablemente a que los hallazgos "interesantes" y exagerados de los estudios no reproducibles atraen más atención, subvenciones y cobertura mediática.

Este hallazgo demuestra un colapso total en la función de señalización del sistema académico, donde las métricas de éxito (citas) están desvinculadas de la validez científica, o incluso se correlacionan inversamente con ella. Esto es, el sistema actual recompensa el "hype" y la novedad por encima del rigor y la fiabilidad.

Este estado de cosas no es un fallo marginal, sino una característica sistémica. La patología económica del mercado de la publicación académica está inextricablemente ligada a su patología epistémica.

¿Una ciencia más replicable conduciría a una mayor confianza pública en los hallazgos científicos? No hay evidencia clara, pero la doxa sugiere que es probable, ya que la lucha por el prestigio y el avance profesional dentro de un sistema basado en la escasez artificial incentiva la producción de conocimiento que es, en un grado alarmante, falso.

A este respecto, conviene recordar que "*la ciencia no busca verdades absolutas, sino refutar errores*" (Popper, *dixit* [39]), esto es, la ciencia avanza por replicación: que se replique es lo que hace que la ciencia sea ciencia: no es casualidad. Y en la medida que los resultados científicos pueden ser importantes para mejorar la vida de las personas (florecimiento humano), resulta obvio la necesidad

---

[33] https://www.polytechnique-insights.com/en/braincamps/society/what-does-it-mean-to-trust-science/science-can-suffer-from-lack-of-reproducibility-of-results/

[34] https://www.nature.com/articles/533452a

[35] https://www.science.org/doi/10.1126/science.aac4716

[36] https://pubmed.ncbi.nlm.nih.gov/31346273/

[37] https://pmc.ncbi.nlm.nih.gov/articles/PMC4461318/

[38] https://www.science.org/doi/10.1126/sciadv.abd1705

[39] https://www.nationalgeographic.com.es/ciencia/principio-falsabilidad-prueba-oro-para-saber-si-algo-es-ciencia_26003

de saber en qué resultados se puede confiar [40]. Además, si la comunidad científica acepta que ciertos hallazgos de investigación son dudosos e intenta mejorar estas deficiencias, tal vez los escépticos de la ciencia sean menos reacios a aceptar resultados de investigación que sean realmente sólidos

La crisis de la comunicación científica no es solo una cuestión de costes; es una crisis de confianza en la propia empresa científica. Para su resolución quizás se debería abandonar la obsesión por la novedad y abrazar la humildad del método, alejándose lo más posible de la colza de la financiarización.

# II. La Constitución Filosófica: Principios Fundamentales para un Nuevo Ecosistema de Conocimiento Humano

Tras diagnosticar las profundas patologías del sistema actual, esta segunda parte pasa de la crítica a la construcción, en un intento de establecer los principios filosóficos, económicos y éticos que deben servir de cimientos para el diseño de un nuevo ecosistema de comunicación académica.

No parece suficiente con sustituir una tecnología por otra; es imperativo construir el nuevo sistema sobre una base de principios coherente y robusta que garantice que sea más justo, eficiente y propicio para el florecimiento humano y la prosperidad compartida. Esto es, el rediseño de un sistema tan complejo como el académico difícilmente puede lograrse mediante ajustes incrementales. Requiere el establecimiento de un nuevo conjunto de primeros principios: una constitución filosófica y económica que defina el propósito, la estructura de gobierno y el ethos tecnológico del nuevo orden. Para evitar replicar las fallas del pasado, el rediseño debería basarse al menos en una síntesis coherente de tres marcos de pensamiento: el Ordoliberalismo para la gobernanza, la Economía Humanista para el propósito y el Humanismo Digital para la tecnología.

## II.1 El Ordoliberalismo como Marco de Gobernanza

El ordoliberalismo [41], desarrollado en la Alemania de la posguerra por la Escuela de Friburgo, constituye una poderosa "tercera vía" entre el laissez-faire clásico y la planificación centralizada. Su premisa fundamental no es la dirección estatal de la economía, sino la construcción de un Estado fuerte que establezca y mantenga un marco jurídico-institucional robusto —una *Wirtschaftsverfassung* o "constitución económica"— destinado a garantizar un orden competitivo.

En este contexto, el Estado no actúa como un agente económico directo, sino como una *Marktpolizei* o "policía del mercado", cuya función esencial es impedir la concentración del poder económico (*Vermachtung*), ya que esta, de no controlarse, erosionaría tanto la competencia como la libertad individual [42]. La interpretación de este principio en clave contemporánea, especialmente en el terreno del derecho de la competencia, revela una profunda vigencia del pensamiento ordoliberal, como muestra la literatura jurídica reciente sobre poder digital y plataformas [43], [44].

La Escuela de Friburgo distinguió cuidadosamente entre dos tipos de política: la *Ordnungspolitik* (política de orden) y la *Prozesspolitik* (política de proceso). La primera diseña y salvaguarda las "reglas del juego" económico; la segunda, en cambio, interviene directamente en el proceso

---

[40] https://pmc.ncbi.nlm.nih.gov/articles/PMC9963456/

[41] https://en.wikipedia.org/wiki/Ordoliberalism

[42]

https://www.researchgate.net/publication/228286927_German_Ordnungstheorie_From_the_Perspective_of_the_New_Institutional_Economics

[43] https://www.scup.com/doi/full/10.5617/oslaw2568

[44] https://www.econstor.eu/bitstream/10419/292595/1/schm.141.3.149.pdf

económico. Para los ordoliberales, el principio rector era inequívoco: planificación estatal de las formas, sí; planificación y control del proceso económico, no.

Esta distinción, de raíz ética y normativa, separaba claramente la creación de un marco institucional estable del intervencionismo coyuntural [45]. Desde esta óptica, el sistema contemporáneo de comunicación científica refleja una notable ausencia de *Ordnungspolitik* efectiva. La concentración en el oligopolio editorial internacional es una forma moderna de *Vermachtung*, donde pocas corporaciones dominan la producción y circulación del conocimiento, configurando una estructura anticompetitiva y dependiente [46].

Frente a ello, la solución no recae en una *Prozesspolitik* científica —una ciencia planificada por el Estado—, sino en el establecimiento de una nueva *Ordnungspolitik* aplicable al "mercado" académico. Esta "constitución para la ciencia" podría articularse mediante arquitecturas institucionales descentralizadas y tecnológicamente mediadas, en las que los contratos inteligentes y las Organizaciones Autónomas Descentralizadas (DAOs) funcionen como árbitros neutrales, asegurando el cumplimiento de las reglas sin injerir ni en el contenido de la investigación ni en el proceso epistemológico mismo [47].

En esta línea, la ordenación digital de la competencia —lo que algunos autores han llamado *Ordoliberalism 2.0*— propone una gobernanza que combina el principio de libertad de acceso con la normatividad codificada en protocolos de transparencia algorítmica [48].

La reinterpretación de la *Ordnungspolitik* en la era digital comporta un desafío ético y político significativo: cómo preservar la libertad institucional de la ciencia y, simultáneamente, impedir nuevas formas de concentración de poder económico o epistémico en el ámbito digital [49]. Así, la regulación descentralizada basada en principios ordoliberales podría ofrecer una alternativa viable al centralismo tecnocrático y al laissez-faire de las plataformas privadas globales, reestableciendo la función del Estado como garante del orden y de la libertad, incluso en escenarios gobernados por algoritmos [50].

## II.2 El Telos Normativo: La Economía Humanista

Un marco de gobernanza, por muy elegante que sea, carece de dirección sin un objetivo normativo claro. El marco ordoliberal, si bien es necesario para crear un mercado funcional, dista mucho de ser es suficiente: proporciona el "cómo" (un orden competitivo), pero no el "porqué". El propósito último del sistema académico no puede ser simplemente la eficiencia económica o la maximización de la producción de artículos. El propósito normativo del nuevo sistema debería definirse explícitamente para garantizar que la eficiencia y la competencia sirvan a fines humanos más elevados. Debe estar orientado al florecimiento humano. La Economía Humanista ofrece este fundamento ético, a modo de brújula normativa.

---

[45] https://journals.openedition.org/oeconomia/690

[46] https://www.tandfonline.com/doi/full/10.1080/09644016.2024.2317108

[47] https://www.researchgate.net/publication/391368625_Future-proofing_the_EU_ordoliberal_governance_and_algorithmic_regulation

[48] https://mcec.umaine.edu/2020/05/26/ordoliberalism-2-0

[49] https://link.springer.com/content/pdf/10.1007/s43681-023-00367-5.pdf

50
https://www.researchgate.net/publication/365101546_The_Making_and_Unmaking_of_Ordoliberal_Language_A_Digital_Conceptual_History_of_European_Competition_Law

## *II.2.1 La Escuela de Economía Humanista.*

La Escuela de Economía Humanista de Barcelona es un movimiento intelectual y científico, que propone una profunda renovación de la ciencia económica mediante la incorporación de la subjetividad humana, la incertidumbre y la complejidad en el análisis económico [51]. La creación oficial tuvo lugar en 2021, cuando la Junta General de la Real Academia de Ciencias Económicas y Financieras de España (RACEF) acordó por unanimidad la aprobación de iniciativas para potenciar este movimiento intelectual y registrando tal denominación en la Oficina Española de Patentes y Marcas.

En sus orígenes intelectuales, la Escuela adoptó y desarrolló la teoría de conjuntos borrosos introducida por Lotfi Zadeh, permitiendo modelar realidades imprecisas y subjetivas. Como señalan Arnold Kaufmann y Jaime Gil Aluja (1986): *"La teoría de los subconjuntos borrosos, lógica difusa o borrosa, es una parte de las matemáticas que se halla perfectamente adaptada al tratamiento tanto de lo subjetivo como de lo incierto. Es un intento de recoger un fenómeno tal cual se presenta en la vida real y realizar su tratamiento sin intentar deformarlo para hacerlo preciso y cierto"* [52].

Así, surgió el pilar conceptual central de dicha Escuela: el "Principio de Simultaneidad Gradual" [53], enunciado por el Dr. Gil Aluja en el Congreso SIGEF de Buenos Aires (1996): *"Toda proposición es a la vez verdadera y falsa, a condición de asignar un grado o nivel a su verdad y un grado o nivel a su falsedad"*.

Este principio rompe con la lógica binaria aristotélica (tercio excluso) y permite incorporar la subjetividad en la decisión económica. Precisamente, al pasar de la binariedad a la multivalencia, se hace posible incorporar la subjetividad en la decisión económica en entornos de incertidumbre: la binariedad resulta cómoda para la formalización de los procesos, pero en cambio convierte la representación formalizada en un relato alejado de la realidad [54].

El otro un pilar filosófico y práctico fundamental de la Escuela de Economía Humanista de Barcelona es el concepto de "Prosperidad Compartida", desarrollado especialmente por el Dr. Gil Aluja en el contexto de su visión de una economía centrada en el bienestar humano colectivo, la justicia social y la reducción de desigualdades. Así, la prosperidad compartida se fundamenta en una premisa inequívoca: "*Una sociedad próspera no lo es cuando no hace próspera a la mayoría de sus ciudadanos*"[55]. Esta máxima sintetiza la idea central de que el progreso económico carece de sentido si no redunda en el bienestar de todos, no solo de unos pocos.

Refinando la idea anterior, en su discurso sobre "El reto de la prosperidad compartida. Papel de las tres culturas ante el siglo XXI" (Sevilla, 2018), el Dr. Gil Aluja argumentó: "*No creceremos más si no repartimos también más y mejor el crecimiento generado. Si así lo hacemos nuestro fruto será el progreso y la reducción de las desigualdades que lastran nuestro crecimiento*". En otras palabras, "*si no sirve al bienestar de todos, la Economía pierde su razón de ser y se convierte en mera especulación peregrina*" [56]. Este "todos" ha sido entendido desde los inicios de la Escuela de Barcelona como universal, trascendiendo fronteras nacionales y culturales [57].

---

[51] https://racef.es/archivos/publicaciones/me58_19_web_racef_libro.pdf

[52] https://www.fuzzyeconomics.com/pdf/01%20borrosos.pdf (p. 18)

[53] http://www.encuentros-multidisciplinares.org/Revistan%C2%BA6/Jaime%20Gil%20Aluja%201.pdf

[54] https://link.springer.com/book/10.1007/978-1-4757-3011-1

[55] https://racef.es/archivos/publicaciones/web_racef_50aniversariogilalujadeff_2.pdf

[56] https://racef.es/archivos/publicaciones/web_racef_sevilla_ms_53_18.pdf

[57] A modo de anécdota de lo interiorizado de la implantación del principio filosófico imperante, la nota de felicitación navideña de 2024 de la RACEF articulaba la siguiente oración: "Que la paz se instale en todos los pueblos y se abran las puertas a una mayor prosperidad compartida" (https://racef.es/es/node/6098)

Por tanto, esta visión rechaza el modelo de crecimiento concentrado que privatiza las ganancias y socializa las pérdidas y, en contra, aboga por una redistribución efectiva como condición *sine qua non* para un desarrollo económico centrado en las personas y amigable con la naturaleza.

## *II.2.2 El Enfoque de las Capacidades*

El Enfoque de las Capacidades de Amartya Sen [58] constituye una de las críticas más relevantes y sofisticadas a las teorías tradicionales del bienestar, como el utilitarismo, centrado en la felicidad o la satisfacción de los deseos individuales, y el recursismo, que privilegia los ingresos o los recursos materiales como indicadores primarios de desarrollo y bienestar. Según Sen, dichas métricas resultan inadecuadas como indicadores del bienestar real, ya que no capturan la diversidad en las condiciones de vida ni las verdaderas oportunidades de las personas. Por ejemplo, el utilitarismo puede legitimar la complacencia ante la opresión mediante la noción de "preferencias adaptativas": los individuos, al enfrentarse a circunstancias restrictivas, pueden aprender a desear menos, manifestando felicidad incluso en contextos de privación severa [59]. De modo análogo, el recursismo ignora que los individuos poseen capacidades desiguales para convertir los mismos recursos en logros valiosos, debido a factores personales, sociales y ambientales [60].

El sistema de comunicación académica contemporáneo presenta dinámicas análogas a las criticadas por Sen. Está dominado por métricas cuantitativas y utilitaristas, como el número de citas, el factor de impacto de las revistas y las cuantías de financiación, que, al igual que el utilitarismo y el recursismo, son indicadores imperfectos del valor científico y social producido. Por ejemplo, el énfasis en tales métricas puede incentivar la persecución de tendencias financiables, en detrimento de preguntas científicas de mayor relevancia social o epistémica, pero menor rentabilidad inmediata [61]. Asimismo, la mera posesión de una subvención o la publicación en revistas de alto impacto no garantizan la calidad o el impacto significativo de la investigación habilitada.

Desde la perspectiva del Enfoque de las Capacidades, se propone una redefinición radical del bienestar y el desarrollo. El bienestar no debe entenderse ni en términos utilitaristas (felicidad subjetiva o satisfacción de deseos) ni recursistas (posesión de recursos), sino como la expansión de las "capacidades" humanas: las libertades sustantivas que las personas efectivamente tienen para alcanzar los tipos de vida que razonablemente valoran. Así, este planteamiento se estructura sobre dos conceptos centrales:

1. Los "funcionamientos" (functionings): se refieren a los distintos "seres y haceres" que una persona puede alcanzar, tales como estar sano, ejercer un trabajo significativo o participar activamente en la vida social y comunitaria. Trasladado al ámbito investigador, los funcionamientos incluirían la realización de experimentos rigurosos, la colaboración abierta, la revisión por pares reflexiva, la mentoría de estudiantes, la comunicación clara de hallazgos y la traducción del conocimiento en beneficios sociales tangibles [62].
2. Las "capacidades" (capabilities): denotan el conjunto real de combinaciones alternativas de funcionamientos que una persona puede efectivamente lograr, es decir, las oportunidades sustantivas para alcanzar funcionamientos valiosos. Bajo este prisma, el objetivo de una política o de un sistema "justo" no es solo proporcionar recursos o maximizar la utilidad, sino ampliar el conjunto real de capacidades de los individuos. Para la ciencia, estas capacidades incluyen la libertad de crear y validar conocimiento fiable, colaborar sin restricciones de intermediarios, acceder libremente al cuerpo completo del saber humano y fomentar la habilidad de la sociedad para aplicar ese conocimiento a la resolución de problemas apremiantes, promoviendo así una "prosperidad compartida" [63], [64].

---

[58] https://kuangaliablog.wordpress.com/wp-content/uploads/2017/07/amartya_kumar_sen_development_as_freedombookfi.pdf

[59] https://iep.utm.edu/sen-cap/

[60] https://www.openbookpublishers.com/books/10.11647/obp.0130

[61] https://doi.org/10.3152/147154305781779588

[62] https://apps.ufs.ac.za/media/dl/userfiles/documents/news/2012_12/2012_12_10_martha_nussbaum_ufs_december_2012.pdf

Además, un valor fundamental en este enfoque es la libertad de elección como componente intrínseco del bienestar. Sen expone el célebre ejemplo de la diferencia entre una persona que ayuna por decisión propia y otra que pasa hambre por carencia de alimentos: aunque el funcionamiento observado es similar (privación alimentaria), el bienestar de ambas es radicalmente distinto, porque la autolimitación voluntaria implica libertad real, mientras que la privación forzada es signo de ausencia de oportunidades [65]. Así, la pobreza se concibe no como simple insuficiencia de ingresos, sino como una "privación de capacidades" [66].

Por tanto, en su aplicación a la comunicación científica, el Enfoque de las Capacidades redefine también su propósito fundamental: el objetivo prioritario del sistema científico no debería ser maximizar métricas simples de salida (como el número de artículos publicados, el factor de impacto o los ingresos por patentes), sino potenciar la ampliación de capacidades críticas tanto para los investigadores como para la sociedad en general [67]. El éxito del sistema radica así en crear un entorno que maximice el potencial de los científicos para llevar a cabo vidas científicas plenas y valiosas, promoviendo tanto la excelencia epistémica como el bienestar social y personal [68].

Este nuevo marco implica medir el éxito institucional no por la producción de resultados fácilmente cuantificables sino por su capacidad para posibilitar prácticas científicas valiosas, tales como la replicación, la publicación de resultados negativos, el acceso y la reutilización de datos, la interdisciplinariedad y la investigación situada en contextos socialmente relevantes [69].

Un sistema que produce artículos de alto impacto, pero bajo una cultura coercitiva de "publicar o perecer", es menos valioso que uno en el que las investigadoras e investigadores eligen libremente perseguir interrogantes arriesgados, innovadores y de largo plazo, con repercusión potencial tanto dentro como fuera de la academia [70], [71]. De ahí que el reto central resida en maximizar la libertad de oportunidad de los investigadores para explorar diversas y valiosas combinaciones de "funcionamientos" científicos [72].

## II.3 El Ethos Tecnológico: El Humanismo Digital como Declaración de Derechos

La construcción de un nuevo orden de la comunicación científica en la era digital exige la adopción de un ethos tecnológico explícito que oriente el diseño de sus herramientas y arquitecturas. El Manifiesto de Viena sobre el Humanismo Digital proporciona estos principios éticos fundamentales, con un postulado central: "*Debemos dar forma a las tecnologías de acuerdo con los valores y las*

---

*necesidades humanas, en lugar de permitir que las tecnologías den forma a los humanos*" [73], [74]. Este principio invierte la relación tradicional entre quienes crean y quienes son transformados por la tecnología, desplazando el eje del mero desarrollo técnico a una agenda centrada en derechos y valores humanos [75].

El manifiesto aboga por un conjunto de principios que deben integrarse en la arquitectura de cualquier sistema digital de importancia social, a saber:

- Diseño para la Democracia y la Inclusión: Las tecnologías deben diseñarse activamente para promover la participación democrática y superar las desigualdades existentes.
- Regulación Efectiva y Antimonopolio: Es necesario establecer regulaciones claras y aplicarlas para restaurar la competitividad y evitar la concentración de poder en monopolios tecnológicos. Los gobiernos no deben dejar todas las decisiones a los mercados.
- Responsabilidad Humana: Las decisiones con consecuencias significativas para los derechos humanos, individuales o colectivos, deben seguir siendo tomadas por humanos, que son responsables y deben rendir cuentas por ellas. Los sistemas automatizados deben apoyar, no reemplazar, la toma de decisiones humana.
- Colaboración Interdisciplinaria: La resolución de los complejos desafíos de la era digital requiere la colaboración entre las disciplinas tecnológicas y las ciencias sociales y las humanidades.

De manera crucial, el Humanismo Digital rechaza explícitamente los "regímenes pseudo-legales basados en términos de servicio privados y contratos opacos de 'click-through'" [76]. Esta es una crítica directa al modelo de negocio de las editoriales académicas y las plataformas tecnológicas, que imponen sus reglas de forma unilateral. También objeta del tecno-solucionismo ingenuo, reconociendo que las posibles soluciones basadas en tecnologías descentralizadas, como la Ciencia Descentralizada (objeto de atención en la parte siguiente de este trabajo), podrían replicar viejas estructuras de poder en nuevas formas digitales si no se someten a un marco ético deliberado. Por el contrario, Humanismo Digital aboga por el diseño activo de las tecnologías para la democracia y la inclusión, la responsabilidad humana indelegable en la toma de decisiones y la regulación efectiva para prevenir la concentración de poder y los monopolios digitales [77].

Por tanto, este ethos actúa como la "declaración de derechos" (las barandillas éticas) para la creación de un nuevo orden de comunicación de la mancomunidad científica. En concreto, exige que la arquitectura de la Ciencia Descentralizada (DeSci) se optimice explícitamente para los valores humanistas de apertura, equidad y gobernanza democrática.

## II.4 Interrelación entre prosperidad compartida y humanismo digital: Hacia la sociedad 5.0

La interrelación entre el concepto de prosperidad compartida de la Economía Humanista y el Humanismo Digital se materializa en una visión convergente: la tecnología debe estar al servicio del bienestar (florecimiento) humano colectivo, no de la concentración de poder o riqueza. Ambos conceptos comparten principios éticos, objetivos redistributivos y una visión crítica del desarrollo tecnológico deshumanizado [78], a saber:

---

[73] https://caiml.org/dighum/dighum-manifesto/

[74] https://galileocommission.org/vienna-manifesto-on-digital-humanism/

[75] https://doi.org/10.1007/978-3-031-86905-1

[76] https://dighum.wien/key-statements-and-conclusions/

[77] https://doi.org/10.1007/978-3-031-12482-2

[78] https://racef.es/es/node/5793

- Centralidad del ser humano. Tanto la prosperidad compartida como el humanismo digital colocan al ser humano en el centro de sus respectivos paradigmas. El humanismo digital sostiene que "*la tecnología debe utilizarse como un medio para promover el bienestar y la realización humana, en lugar de simplemente como un fin en sí mismo*". Esta máxima es análoga a la visión del Dr. Gil Aluja: "*Si no sirve al bienestar de todos, la Economía pierde su razón de ser*" [79].
- Justicia social y equidad. El Humanismo Digital aboga por una sociedad "*más justa e inclusiva*" [80], donde todas las personas merecen ser tratadas con dignidad y respeto, independientemente de su raza, género, orientación sexual, religión, etc. Este principio se alinea perfectamente con la visión de prosperidad compartida, que exige reducción de desigualdades, inclusión y distribución equitativa del crecimiento económico.
- Inclusión y diversidad. Ambos paradigmas enfatizan la inclusión universal. El Humanismo Digital promueve que la tecnología y la conectividad deben utilizarse de manera que beneficien a todas las personas, independientemente de su edad, género, raza, etnia, discapacidad o ubicación geográfica [81]. Paralelamente, como se mencionó, el Dr. Gil Aluja insiste en que "*no creceremos más si no repartimos también más y mejor el crecimiento generado*" y que la diversidad de talentos es fundamental para la prosperidad compartida

En definitiva, la visión conjunta de la Economía Humanista y el Humanismo Digital obliga a evaluar cada mecanismo y cada protocolo en la era digital con una pregunta dual: ¿expande o restringe las capacidades humanas clave, o en otras palabras, el sistema se optimiza para el florecimiento humano y la prosperidad compartida o para la engrandecer la financiarización particular (especulación financiera y cortoplacismo) sin miramiento alguno de sostenibilidad o inclusividad?

Finalmente, no se debería soslayar que la interrelación entre prosperidad compartida y humanismo digital encuentra eco en la denominada Sociedad 5.0, "*una sociedad centrada en el ser humano que equilibra el avance económico y tecnológico para resolver los problemas de la sociedad con sistemas de datos superinteligentes. Representa una nueva visión para una sociedad más inteligente, donde los seres humanos, la naturaleza y la tecnología crean un equilibrio sostenible mejorado por los datos*" [82].

El concepto de Sociedad 5.0 fue propuesto por primera vez en 2015, en el Quinto Plan Básico de Ciencia y Tecnología por el gobierno japonés como una sociedad futura a la que Japón debería aspirar y que fue anunciado al mundo por el primer ministro japonés Shinzo Abe en su discurso "Declaración de Hannover" en la conferencia CeBIT 2017 [83].

Desde entonces, los mismos principios operativos han sido compartidos por la comunidad internacional. Es el caso de la Declaración Europea sobre los Derechos y Principios Digitales para la Década Digital [84] (ver Figura 1).

En el mismo sentido cabe señalar la "Declaración sobre un futuro digital fiable, sostenible e inclusivo" de la OCDE [85] y también con la iniciativa del Presidente Yoon Suk Yeol de Corea del Sur de "Carta sobre Valores y Principios para una Sociedad Digital de Prosperidad Compartida" [86] que resuenan perfectamente con la visión de la Escuela Humanista propugnada por la RACEF, a saber:

---

[79] https://racef.es/archivos/publicaciones/web_racef_50aniversariogilalujadeff_2.pdf (p. 10)

[80] https://fundacionhermes.org/una-perspectiva-humanista-al-servicio-del-progreso-digital-para-la-mayoria/

[81] https://www.emancipatic.org/decalogo-humanismo-digital-personas-mayores/

[82] https://buleria.unileon.es/bitstream/handle/10612/26358/Argocapitalismo_Multinter.net%20_%20Hacia_Nuevas_Competencias_Trabajo_Futuro%20.pdf?sequence=1&isAllowed=y

[83] https://www.japan.go.jp/letters/ebook56/book.pdf

[84] https://eur-lex.europa.eu/legal-content/ES/TXT/HTML/?uri=CELEX:32023C0123(01)

[85] https://legalinstruments.oecd.org/api/download/?uri=/public/f347ae7c-4e38-4e7a-8d29-af489cffb84e.pdf

[86] https://www.korea.net/Government/Briefing-Room/Press-Releases/view?articleId=7042&type=O

- Garantía de libertad y derechos: "*El fundamento de la sociedad digital debe descansar en el respeto por la dignidad y los valores humanos, asegurando que cada individuo tenga garantizadas la libertad y los derechos en el entorno digital*".
- Acceso justo y oportunidades equitativas: "*Las oportunidades de competición e innovación deben asegurarse de manera justa para todos, y los beneficios de la innovación digital deben distribuirse equitativamente en la comunidad*".
- Promoción del bienestar humano: "*Las naciones deben colaborar con la comunidad internacional, guiadas por valores universales y confianza mutua, para aprovechar la tecnología digital para la mejora del bienestar humano y la reducción de la brecha digital entre naciones*".

**Figura 1. Declaración Europea sobre los Derechos y Principios Digitales para la Década Digital**
Fuente: https://ec.europa.eu/newsroom/dae/redirection/document/94370

# III. La Arquitectura de un Ecosistema Académico Descentralizado

Esta tercera parte traduce los principios rectores de la Parte II en un plan arquitectónico concreto y factible. Este plan debería traducir la constitución ordoliberal y humanista en una infraestructura técnica e institucional operativa.
La Ciencia Descentralizada (DeSci) proporciona el conjunto de herramientas tecnológicas necesarias para construir este nuevo orden epistémico [87]. Así, en esta parte se detalla un ecosistema de extremo a extremo para el ciclo de vida de la investigación, desde la financiación y la ejecución hasta la revisión por pares, la publicación y la gestión de la reputación.

---

[87] https://storage.prod.researchhub.com/uploads/papers/2024/01/27/preprints202401.1638.v1.pdf

La arquitectura propuesta se basa en cuatro pilares interconectados: la gobernanza a través de Comunidades de Investigación Autónomas (ARCs), la propiedad intelectual como activos líquidos y componibles (IP-NFTs), la reputación como credencial inmutable (Tokens Soulbound) y la financiación como un bien público pluralista (Financiación Cuadrática). Juntos, estos componentes forman un sistema auto-reforzante diseñado para realinear los incentivos con los objetivos fundamentales de la ciencia.

Para hacer frente a un problema sistémico, se necesita una solución sistémica: la DeSci emerge como una alternativa fundamental que no busca reformar el modelo editorial existente, sino reemplazarlo. Este no es un futuro especulativo, sino un sistema que puede construirse con las tecnologías existentes, guiado por una elección institucional deliberada.

A modo de resumen introductorio, la Tabla 1 muestra las principales diferencias que caracterizan el modelo centralizado versus la propuesta descentralizada, cuyo detalle se amplía en los siguientes epígrafes.

**Tabla 1: Un Marco Comparativo de Paradigmas Académicos**

| Característica | Paradigma Centralizado Actual | Paradigma Descentralizado Propuesto |
|---|---|---|
| **Gobernanza** | Consejos editoriales opacos; administración universitaria centralizada; agencias de subvenciones gubernamentales. | Comunidades de Investigación Autónomas (ARCs) transparentes, gobernadas por sus miembros mediante votación híbrida de reputación/cuadrática. |
| **Financiación** | Procesos de subvención centralizados, lentos y sesgados; capital de riesgo para la PI en etapas tardías. | Financiación Cuadrática democrática para la investigación como bien público; mercados líquidos y globales para la PI en etapas tempranas a través de IP-NFTs. |
| **Revisión y Validación** | Revisión pre-publicación anónima, no remunerada, lenta e irresponsable. Devaluación sistémica de la replicación. | Revisión post-publicación pública, incentivada (con tokens) y atribuible. Creación de reputación (vía SBTs) por revisiones rigurosas y replicaciones exitosas. |
| **Propiedad Intelectual** | Ilíquida, aislada en oficinas de transferencia de tecnología; empaquetada en patentes monolíticas. | IP-NFTs líquidos, componibles y de propiedad fraccionable, que permiten nuevos modelos de propiedad y financiación colectiva. |
| **Reputación y Credenciales** | Indicadores indirectos de calidad: Factor de Impacto de la revista, índice h, afiliación institucional. | Contribuciones verificables, no transferibles y *on-chain* a través de Tokens Soulbound (SBTs) para acciones específicas (revisión, datos compartidos, replicación). |

| **Acceso y Difusión** | Acceso restringido por muros de pago controlados por editoriales oligopólicas; largos retrasos en la publicación. | Acceso abierto, permanente, incensurable a través de almacenamiento descentralizado (IPFS/Arweave); publicación inmediata de *preprints*. |
|---|---|---|
| **Lógica Económica Central** | Escasez y extracción de prestigio (búsqueda de rentas). | Abundancia, colaboración y creación de valor (generativa). |

## III.1 El Stack DeSci como el Nuevo Locus de la Producción de Conocimiento

La Ciencia Descentralizada (DeSci) no es una única tecnología, sino una pila o "stack" (arquitectura en capas) que aprovecha la tecnología blockchain para reimaginar el proceso científico [88], [89]. En su nivel más fundamental, la blockchain proporciona una capa de registro segura, transparente e inmutable, un libro de contabilidad público que puede registrar transacciones, datos y acuerdos de forma verificable y a prueba de manipulaciones [90]. Sobre esta base, los contratos inteligentes (*smart contracts*) actúan como código autoejecutable que automatiza las reglas y los acuerdos, eliminando la necesidad de intermediarios para hacer cumplir los contratos. En la capa superior, las Organizaciones Autónomas Descentralizadas (DAOs) ofrecen un marco para la gobernanza comunitaria, permitiendo a los participantes tomar decisiones colectivas sobre los recursos y las reglas del sistema [91], [92].

El objetivo de este stack tecnológico es reemplazar las estructuras cerradas y burocráticas de la ciencia tradicional —revistas con muros de pago, comités de subvenciones opacos— por una economía del conocimiento transparente y gobernada mediante programación [93].

El ecosistema DeSci ya está desarrollando plataformas para la publicación descentralizada, la financiación colectiva, el almacenamiento de datos resistente a la censura y la colaboración abierta [94]. Este conjunto de herramientas es lo que hace factible la implementación de una *Wirtschaftsverfassung* ordoliberal en el siglo XXI. La "constitución económica" de la ciencia ya no es un documento estático, sino un conjunto de contratos inteligentes interoperables y autoejecutables que definen las reglas del juego para la producción de conocimiento.

## III.2 Gobernanza por Comunidades de Investigación Autónomas (ARCs)

Como se mencionó en el epígrafe anterior, las DAOs son el bloque de construcción institucional central del nuevo ecosistema. Funcionan como sociedades científicas del siglo XXI, nativas digitalmente, organizadas en torno a disciplinas, subdisciplinas o misiones de investigación específicas [95], [96], [97]. Sus responsabilidades principales incluirían la gestión de procesos de revisión por pares transparentes y potencialmente incentivados, la gobernanza de tesorerías gestionadas por la comunidad para la financiación de la investigación y el establecimiento de estándares de datos, ética y reproducibilidad dentro de su dominio.

---

[88] https://www.researchgate.net/publication/382370008_Decentralized_science_DeSci_definition_shared_values_and_guiding_principles

[89] https://chain.link/education-hub/decentralized-science-desci

[90] https://hacken.io/discover/blockchain-architecture-layers/

[91] https://www.rapidinnovation.io/post/the-role-of-daos-and-blockchain

[92] https://dev.to/swarmzero/understanding-the-decentralized-ai-stack-5e3o

[93] https://coinpaper.com/9693/de-sci-explained-how-science-without-borders-works

[94] https://99bitcoins.com/education/what-is-desci/

## *III.2.1 Afrontar los Fallos de Gobernanza*

En el corazón del nuevo sistema se encuentran las Comunidades de Investigación Autónomas (ARCs), que son DAOs especializadas en dominios científicos específicos. Estas ARCs funcionan como los órganos de gobierno descentralizados que supervisan la revisión por pares, la validación de la investigación y la asignación de recursos dentro de sus campos correspondientes, pero en sus actuaciones no se libran de presentar distintos problemas sistémicos.

Resulta crucial, por tanto, abordar directamente los modos de fallo bien documentados de los modelos simplistas de DAOs. El modelo más común, el modelo de gobernanza estándar en el mundo de las DAOs, "*un token, un voto*", es fundamentalmente defectuoso para la ciencia, ya que es inherentemente plutocrático [98]. Este modelo crea el "*Problema de la Plutocracia en las DAOs*", donde la influencia en la toma de decisiones se deriva de la riqueza (el número de tokens que se poseen) en lugar de la experiencia o el mérito [99], [100]. Esto es, un pequeño número de grandes poseedores de tokens ("ballenas") pueden capturar la gobernanza y dirigir las decisiones en su propio interés, contradiciendo el espíritu descentralizado [101], [102], [103].

Otro reto importante es la apatía de los votantes, donde las bajas tasas de participación llevan a que las decisiones sean tomadas por una pequeña minoría activa, socavando la legitimidad del proceso [104].

Por último, los ataques de gobernanza, en los que actores maliciosos utilizan tácticas insidiosas para adquirir temporalmente un gran poder de voto y aprobar propuestas que les permitan robar fondos, representan una amenaza de seguridad significativa [105], [106], [107]. Esta vulnerabilidad no es teórica; ya ha sido explotada en el mundo real, como en el caso de *Beanstalk DAO*, donde un atacante utilizó un "préstamo relámpago" (*flash loan*) para adquirir temporalmente una enorme cantidad de tokens de gobernanza, aprobar una propuesta maliciosa y drenar 182 millones de dólares de la tesorería [108]. Por

---

[95] https://www.scielo.br/j/bar/a/nqdm7FjmhVTyg9bV4WPJLcC/?lang=en

[96] https://www.rapidinnovation.io/post/daos-explained-ultimate-guide-to-decentralized-autonomous-organizations

[97] https://www.sciencedirect.com/science/article/pii/S0304405X25000011

[98] https://serto.medium.com/the-dao-plutocracy-problem-a8841546a0f2

[99] https://medium.com/mosaic-network-blog/is-plutocratic-on-chain-governance-really-a-bad-thing-68132700205c

[100] https://www.belfercenter.org/publication/deep-dive-citydao-experiment-collective-land-ownership-and-decentralized-governance

[101] https://prism.sustainability-directory.com/scenario/climate-dao-governance-best-practices/

[102] https://www.coinfabrik.com/blog/who-really-rules-web3/

[103] https://www.researchgate.net/publication/381214311_Analyzing_Voting_Power_in_Decentralized_Governance_Who_controls_DAOs

[104] https://arxiv.org/abs/2407.21461

[105] https://www.cyfrin.io/glossary/governance-attack

[106] https://www.quillaudits.com/blog/web3-security/dao-governance-attacks

[107] https://en.wikipedia.org/wiki/The_DAO

[108] https://nhsjs.com/2025/strengthening-dao-governance-vulnerabilities-and-solutions/

tanto, un sistema científico gobernado de esta forma sería susceptible de ser capturado por intereses puramente financieros.

## *III.2.2 Modelos de Gobernanza Avanzados como Solución*

Para mitigar estos riesgos, la arquitectura propuesta avanza hacia modelos de "DAO 3.0" más sofisticados y resistentes (más allá de la simple votación por tokens), que equilibran la automatización técnica con la complejidad humana y la confianza relacional [109]. La elección de este modelo no es una sutileza técnica, sino una necesidad filosófica para hacer realidad los principios de la Parte II, pues un modelo de DAO ingenuo basado en tokens violaría los principios del Humanismo Digital (anti-plutocracia) y de la Economía Humanista (privilegiando los recursos sobre los funcionamientos obviando la prosperidad compartida).

De esta forma, se aboga nítidamente en que el poder de gobierno dentro de las ARCs no se base únicamente en la tenencia de tokens. En su lugar, se sugiere implementar un modelo híbrido que combine dos mecanismos avanzados:

- Gobernanza Basada en la Reputación: La influencia en la votación se pondera según la reputación no transferible de un usuario, que se gana a través de contribuciones positivas y verificables a la comunidad, como la realización de revisiones por pares de alta calidad o la publicación de datos replicables [110]. Esto alinea el poder de decisión con el mérito intelectual demostrado y garantiza que la influencia se correlacione con la contribución científica, no con la riqueza, rompiendo deliberadamente el vínculo entre el poder financiero y el poder de gobierno [111], [112].
- Votación Cuadrática: Este mecanismo permite a los participantes expresar la intensidad de sus preferencias de una forma que mitiga el poder de las "ballenas" (grandes poseedores de tokens). En este sistema, el coste de votos adicionales aumenta cuadráticamente: mientras la primera votación de un individuo cuesta una unidad, la segunda cuesta cuatro, la tercera nueve, y así sucesivamente. Esto amplifica las voces de muchos pequeños interesados sobre la de unos pocos grandes, fomentando una toma de decisiones más democrática [113], [114], ya que da más peso a la amplitud del apoyo (el número de personas que apoyan una propuesta) que a la profundidad del apoyo de unos pocos actores poderosos [115].
- Delegación y democracia líquida: Los miembros pueden delegar su poder de voto en expertos de confianza dentro de la comunidad, creando un proceso de toma de decisiones más escalable y eficiente sin sacrificar la participación [116].
- Mejores prácticas de seguridad: Para prevenir ataques, es esencial implementar bloqueos de tiempo entre las fases de propuesta, votación y ejecución, lo que anula la eficacia de los ataques de préstamo flash. Además, se pueden establecer mecanismos de veto, controlados por un consejo multisignatura de miembros de confianza, para detener propuestas maliciosas en caso de emergencia [117].

---

[109] https://www.frontiersin.org/journals/blockchain/articles/10.3389/fbloc.2025.1630402/full

[110] https://blog.colony.io/understanding-dao-voting-mechanisms-a-focus-on-colony-2/

[111] https://www.sciencedirect.com/science/article/pii/S0304405X25000011

[112] https://papers.ssrn.com/sol3/papers.cfm?abstract_id=4485228

[113] https://www.rapidinnovation.io/post/daos-explained-ultimate-guide-to-decentralized-autonomous-organizations

[114] https://metana.io/blog/dao-governance-models-what-you-need-to-know/

[115] https://journals.sagepub.com/doi/10.1177/20539517241293803

[116] https://www.elsevier.es/en-revista-journal-innovation-knowledge-376-articulo-built-last-not-scale-long-S2444569X24000416

[117] https://www.quillaudits.com/blog/web3-security/dao-governance-attacks

Por tanto, este modelo de gobernanza híbrido aborda directamente la crítica de la plutocracia y se alinea con el principio de inclusión y florecimiento humano. Crea un sistema en el que tanto la contribución de capital (tokens) como la contribución intelectual (reputación) son valoradas, evitando la captura de la ciencia por intereses puramente financieros.

El fracaso histórico de "The DAO" en 2016, que se basó en una gobernanza puramente basada en código sin salvaguardias sociales, sirve como una advertencia permanente sobre los peligros de la simplificación excesiva [118], [119].

## III.3 Financiando la Ciencia como un Bien Público: De las Subvenciones a los Flujos Cuadráticos

El modelo actual de financiación de la investigación, basado en subvenciones centralizadas, es ampliamente criticado por ser lento, burocrático, excesivamente conservador y susceptible a sesgos, lo que a menudo ahoga la innovación en lugar de fomentarla [120], [121]. El nuevo ecosistema propone una arquitectura de financiación híbrida y de dos vías para superar estas limitaciones.

### III.3.1 Financiación Cuadrática (QF) para Bienes Públicos

Para la investigación científica fundamental, impulsada por la curiosidad y la búsqueda de la verdad, que puede no tener una aplicación comercial inmediata, pero que cuenta con un amplio apoyo de la comunidad, la Financiación Cuadrática (QF) ofrece una alternativa radicalmente más democrática y eficiente para financiar bienes públicos [122], [123], [124], [125].

El mecanismo de la QF utiliza un fondo de contrapartida (procedente de fundaciones, gobiernos o tesorerías de DAOs) para amplificar las pequeñas contribuciones de un gran número de personas por encima de las grandes contribuciones de unos pocos. La fórmula matemática subyacente asigna fondos de contrapartida a un proyecto de forma proporcional al cuadrado de la suma de las raíces cuadradas de las contribuciones individuales [126], [127], [128].

En la práctica, esto significa que el número de personas que apoyan un proyecto es mucho más importante que la cantidad total que donan. Un proyecto con 100 donantes de 1 dólar recibirá una contrapartida mucho mayor que un proyecto con un solo donante de 100 dólares. Así, este mecanismo es intrínsecamente democrático y crea una señal de mercado descentralizada y ascendente sobre qué investigación de bien público valora más la comunidad científica [129], [130].

---

En la arquitectura propuesta, las DAOs de financiación a gran escala utilizarán la QF para asignar capital. En lugar de que un pequeño comité de expertos decida sobre subvenciones multimillonarias a puerta cerrada, el fondo de contrapartida se distribuirá según la sabiduría colectiva de toda la comunidad científica, expresada a través de micro-donaciones.

Plataformas como *ResearchHub* ya están experimentando con modelos de tokenómica para incentivar la financiación comunitaria de la investigación, donde los usuarios pueden usar tokens para crear subvenciones y recompensar el trabajo científico [131], [132], [133]. Este modelo alinea la financiación con las prioridades colectivas de la propia comunidad científica, fomentando un ecosistema de investigación más diverso, resiliente y orientado al bien público.

### *III.3.2 DAOs Orientadas a la Misión (El Modelo ARPA-H)*

Para los proyectos de alto riesgo, alta recompensa y "moonshot" que abordan grandes desafíos sociales, el sistema necesita un mecanismo que abrace el fracaso y persiga la innovación disruptiva. Para ello, la arquitectura propuesta se respalda la creación de DAOs orientadas a la misión, modeladas a partir de las experiencias llevadas a cabo por la Agencia de Proyectos de Investigación Avanzada de Defensa (DARPA) y su homóloga en salud, ARPA-H [134], [135].

Estas DAOs estarían dirigidas por "Directores de Programa" capacitados con una autonomía significativa para financiar enfoques diversos y competitivos para resolver problemas ambiciosos y bien definidos. Este modelo, a diferencia de los comités de subvenciones tradicionales, no busca el consenso, sino que fomenta enfoques radicales que las vías de investigación tradicionales o comerciales no apoyarían [136], [137].

Este modelo dual de financiación híbrido implementa directamente la distinción ordoliberal entre un proceso competitivo ascendente y una intervención específica para objetivos estratégicos. La Financiación Cuadrática crea un vibrante "mercado" descentralizado de ideas de investigación, mientras que las DAOs orientadas a la misión actúan como una herramienta estratégica para abordar los principales desafíos sociales, sin planificar centralmente todo el panorama de la investigación. Esto crea un sistema que es a la vez resistente y dinámico, apoyando tanto la investigación impulsada por la curiosidad como los avances específicos.

## III.4 Reclamando la Propiedad Intelectual: IP-NFTs y el Valor del Conocimiento

Uno de los mayores obstáculos en la ciencia es el "valle de la muerte" de la financiación en etapas tempranas, donde las ideas prometedoras a menudo perecen por falta de capital para pasar de la investigación básica a la prueba de concepto. Otro estorbo para la innovación se encuentra en el mismo sistema actual de propiedad intelectual (IP), donde leyes como la Ley Bayh-Dole en los EE. UU., aunque bien intencionadas para promover la comercialización, a menudo han llevado a que las universidades actúen como maximizadoras de ingresos, con oficinas de transferencia de tecnología (OTT) burocráticas que pueden retrasar o impedir la transferencia de tecnología en lugar de facilitarla [138], [139], [140], [141]. El problema fundamental es que todos los derechos se agrupan en un único activo ilíquido controlado por un intermediario lento.

---

[131] https://researchhub.foundation/research-coin

[132] https://blog.researchhub.foundation/the-4-steps-to-get-your-research-funded-by-a-desci-dao-a-step-by-step-guide/

[133] https://docs.researchhub.com/researchcoin/what-is-researchcoin

[134] https://research.jhu.edu/rdt/about-arpa-h/

[135] https://arpa-h.gov/about/arpa-h-model

[136] https://arpa-h.gov/explore-funding/open-funding-opportunities

[137] https://arpa-h.gov/explore-funding/submission-resources-and-FAQs/mission-office-iso-know-before-applying

Para resolver estos impedimentos, el modelo propuesto los aborda transformando la IP de un activo ilíquido y aislado a uno líquido, componible y negociable a nivel mundial, para lo cual la herramienta bien pudiera ser el Token No Fungible de Propiedad Intelectual (IP-NFT) [142] que es un token criptográfico único que representa la propiedad segura, verificable y transferible de un activo intangible —como una patente, un conjunto de datos, un protocolo de investigación o un descubrimiento— en la blockchain [143]. De hecho, protocolos pioneros como *Molecule* ya están permitiendo a las *BioDAOs* (DAOs centradas en la biotecnología) como *VitaDAO* financiar investigaciones a cambio de IP-NFTs que representan la IP resultante [144], [145].

Los IP-NFTs ofrecen una solución al permitir la tokenización y el desglose de los diversos derechos asociados a un producto de investigación, como una patente, un conjunto de datos o un descubrimiento [146], [147]. Al objeto de comprender su funcionamiento, resulta crucial entender la distinción legal entre la propiedad del token (*corpus mechanicum*, la representación digital) y la propiedad de los derechos de IP subyacentes (*corpus mysticum*, el activo intangible), que están codificados en el contrato inteligente del token [148], [149]. Así, la compra de un NFT no transfiere automáticamente los derechos de autor o la marca registrada subyacentes, a menos que se especifique explícitamente en los términos del contrato inteligente [150].

Los IP-NFTs permiten desarrollar nuevos modelos de financiación y comercialización de la investigación que puede democratizar la financiación y acelerar la traslación [151], [152], a saber:

- Propiedad fraccionada: Los proyectos de investigación pueden financiarse vendiendo la propiedad fraccionada de la futura PI a una comunidad de inversores a través de una DAO. Esto abre la financiación de la ciencia a un grupo mucho más amplio de partes interesadas.
- Regalías automatizadas: Los contratos inteligentes pueden distribuir automáticamente los ingresos por licencias a todas las partes interesadas —los investigadores, sus instituciones, la

---

[138] http://www.ipadvocatefoundation.org/forum/topic-751.cfm

[139] https://pmc.ncbi.nlm.nih.gov/articles/PMC11152831/

[140] https://brie.berkeley.edu/sites/default/files/wp182.pdf

[141] https://direct.mit.edu/daed/article/147/4/76/27224/Universities-The-Fallen-Angels-of-Bayh-Dole

[142] https://www.uspto.gov/sites/default/files/documents/Joint-USPTO-USCO-Report-on-NFTs-and-Intellectual-Property.pdf

[143] https://www.researchgate.net/publication/384759619_WHEN_IP_MEETS_NFTs_THE_CASE_FOR_INTELLECTUAL_PROPERTY_PROTECTION_FOR_NON-FUNGIBLE_TOKENS

[144] https://www.gate.com/learn/articles/a-comprehensive-overview-of-molecule/5404

[145] https://molecule.xyz/learn/cryptocurrency-for-scientists

[146] https://en.wikipedia.org/wiki/Non-fungible_token

[147] https://www.emerald.com/medar/article/33/7/385/1267641/Business-model-digitalization-and-decentralization

[148] https://www.wipo.int/en/web/wipo-magazine/articles/the-metaverse-nfts-and-ip-rights-to-regulate-or-not-to-regulate-42603

[149] https://hedera.com/learning/nfts/nft-intellectual-property

[150] https://www.tamimi.com/law-update-articles/nfts-intellectual-property-and-other-legal-considerations-a-wake-up-call/

[151] https://medium.com/1kxnetwork/fungible-non-fungibles-the-financialization-of-nfts-32565adf454a

[152] https://discovery.ucl.ac.uk/10145537/2/Chiu_Rev%20clean%20What%20Purposes%20do%20NFTs%20Serve3.8_JGA.pdf

- DAO de financiación e incluso los contribuyentes ciudadanos— a perpetuidad y sin los gastos generales de una OTT.
- Mercados líquidos: Al crear un mercado transparente y líquido para la PI de la investigación, los IP-NFTs pueden mejorar el descubrimiento de precios y facilitar que los descubrimientos en fase inicial encuentren socios comerciales.

La idea principal radica en que los contratos inteligentes que rigen estos IP-NFTs pueden hacer cumplir programáticamente reglas alineadas con el bien público. Por ejemplo, un IP-NFT para una nueva patente farmacéutica podría incluir una cláusula que obligue a que el artículo de investigación asociado sea permanentemente de acceso abierto, o que los datos de los ensayos clínicos subyacentes se hagan públicos después de un período de embargo. Esto utiliza mecanismos de mercado para lograr objetivos de política pública.

Este modelo resuelve la tensión entre el bien público y la comercialización que ha afectado a la transferencia de tecnología universitaria. Al desglosar los derechos y automatizar los flujos de ingresos, permite tanto el acceso abierto al conocimiento como los incentivos financieros necesarios para el desarrollo comercial. El contrato inteligente puede, simultáneamente, conceder una licencia universal e irrevocable para la investigación y la lectura no comerciales (garantizando el acceso abierto) y regir la venta de licencias exclusivas para la aplicación comercial. Esto alinea los incentivos de los científicos, los financiadores y el público en un único sistema transparente, convirtiendo un juego de suma cero en uno de suma positiva. Aunque la aplicabilidad legal de los derechos codificados en los NFT todavía está evolucionando, los tribunales ya han comenzado a reconocer las infracciones de marcas registradas a través de los NFT, sentando un precedente para el reconocimiento legal de los derechos de IP en la cadena [153], [154].

Por tanto, en la arquitectura propuesta, el flujo de trabajo es el siguiente: un investigador o un laboratorio acuña un IP-NFT para un nuevo descubrimiento. Este IP-NFT se transfiere a la tesorería de una ARC a cambio de financiación para continuar la investigación. La ARC, a su vez, puede fraccionar este IP-NFT en Tokens de Propiedad Intelectual (IPTs) fungibles, que pueden ser distribuidos a los miembros de la comunidad que financiaron el proyecto [155], [156]. Esto permite la propiedad colectiva y la gobernanza sobre la estrategia de comercialización de la IP, con los ingresos de las licencias o ventas fluyendo de vuelta a la tesorería de la DAO para financiar más investigaciones.

## III.5 La Reputación como un Activo Intransferible: Tokens Soulbound (SBTs)

La patología central del sistema actual es su dependencia de indicadores indirectos y fácilmente manipulables de calidad y prestigio, como el factor de impacto de las revistas y el índice h, que son la causa principal de la crisis epistémica. En su lugar, se propone un sistema de reputación multidimensional, granular, verificable y controlado por el usuario, construido sobre la base de los Tokens Vinculados al Alma (Soulbound Tokens, SBTs) [157].

---

[153] https://www.hoganlovells.com/en/publications/trademark-infringement-through-nfts-intellectual-property-enforcement-in-the-virtual-world

[154] https://www.tamimi.com/law-update-articles/nfts-intellectual-property-and-other-legal-considerations-a-wake-up-call/

[155] https://www.molecule.to/blog/ipts-a-gain-of-function

[156] https://www.wipo.int/en/web/wipo-magazine/articles/the-metaverse-nfts-and-ip-rights-to-regulate-or-not-to-regulate-42603

[157] https://www.webopedia.com/crypto/learn/what-is-a-soulbound-token/

Un SBT es un tipo especial de NFT que, una vez emitidos a una dirección de monedero digital, no pueden ser transferidos [158]. Se trata de un NFT no transferible que está permanentemente vinculado a una identidad digital única o "alma" (una cartera de blockchain) [159]. Su no transferibilidad los hace ideales para representar logros y afiliaciones personales que no deben ser comprados ni vendidos, como títulos académicos, certificaciones profesionales o, en este contexto, contribuciones científicas [160].

En el nuevo orden epistémico aquí propugnado, la reputación es la moneda principal. El "Alma" de un investigador acumularía SBTs que representan todo el espectro de sus contribuciones a la mancomunidad científica, creando un currículum vitae en la cadena de bloques, rico y verificable:

- Publicaciones: Un SBT emitido por una DAO de revisión por pares por cada artículo aceptado o por publicar un *preprint* que incluya datos abiertos y código reproducible, validado por la comunidad (a través de plataformas como *DeSci Publish*, que ya permiten agrupar manuscritos, datos y código) [161], [162].
- Revisiones por pares: Un SBT por cada revisión por pares de alta calidad completada, rigurosa y constructiva (haciendo la revisión pública y atribuible), esto es, reconociendo y recompensando esta labor crucial, pero actualmente invisible.
- Replicaciones: SBTs de alto valor por replicar con éxito (o no replicar) estudios importantes, incentivando directamente la validación de la ciencia.
- Compartición de datos y código: SBTs por publicar conjuntos de datos bien documentados, una nueva metodología o una herramienta de software de código abierto.
- Mentoría: SBTs emitidos por los estudiantes tras su graduación o la consecución de logros importantes.
- Credenciales: SBTs de las universidades por los títulos obtenidos, creando un diploma digital a prueba de falsificaciones [163], [164].

Por tanto, este sistema crea los currícula *on-chain* (en la cadena de bloques) verificables, granulares y portátiles. Desacopla la reputación de un investigador de las revistas en las que publica y la vincula directamente a sus contribuciones verificables al bien común científico. Esto incentiva directamente los comportamientos —apertura, colaboración, rigor, replicación— que son esenciales para una ciencia robusta, pero que el sistema actual penaliza o ignora.

No obstante, nuestra propuesta no está exenta de importantes riesgos para la privacidad [165]. Para mitigar estos riesgos, se proponen soluciones técnicas que permiten la divulgación selectiva y la preservación de la privacidad, a saber:

- Almacenamiento fuera de la cadena: Los metadatos sensibles de los SBTs (por ejemplo, los detalles de una revisión por pares confidencial) se almacenarían fuera de la cadena en una red de almacenamiento descentralizada como IPFS o en una bóveda de datos privada. Solo se almacenaría en la cadena un hash o un compromiso criptográfico [166].

---

[158] https://support.exodus.com/support/en/articles/9563015-what-is-a-soulbound-token-sbt

[159] https://support.exodus.com/support/en/articles/9563015-what-is-a-soulbound-token-sbt

[160] https://store.aicerts.ai/blog/the-rise-of-soulbound-tokens-sbts-a-new-frontier-for-non-transferable-nfts/

[161] https://www.desci.com/publish

[162] https://www.desci.com/blog/decentralized-open-access-what-web3-means-for-scientific-publishing?utm_source=social

[163] https://store.aicerts.ai/blog/the-rise-of-soulbound-tokens-sbts-a-new-frontier-for-non-transferable-nfts

[164] https://www.leewayhertz.com/soulbound-tokens/

[165] https://www.webopedia.com/crypto/learn/what-is-a-soulbound-token/

[166] https://www.itm-conferences.org/articles/itmconf/pdf/2023/06/itmconf_icdsac2023_06002.pdf

- Pruebas de conocimiento cero (ZKPs): Esta poderosa técnica criptográfica permitiría a un investigador demostrar hechos sobre su reputación sin revelar los datos subyacentes [167]. Por ejemplo, podrían demostrar que "*he completado 7 revisiones por pares para revistas de finanzas*" sin revelar qué artículos específicos revisaron.

Finalmente, cabe señalar que este sistema de reputación rico, verificable y controlado por el usuario formaría la base de una nueva economía del prestigio: se podría utilizar para la gobernanza de las DAOs, las decisiones de contratación y la identificación de expertos.

Además, este sistema basado en SBTs podría erigirse como el eje que conecta todo el ecosistema rediseñado. Proporciona la entrada no financiera y basada en el mérito necesaria para que la gobernanza de las DAOs funcione, resolviendo así el problema de la plutocracia. Al mismo tiempo, crea incentivos positivos que contrarrestan directamente los impulsores de la crisis de replicación, al recompensar actividades como la revisión por pares de calidad y los estudios de replicación.

Por tanto, el sistema de SBTs más allá de ser un simple sustituto del índice h, está revolucionando la investigación y el desarrollo científicos [168].

## III.6 El Ciclo Virtuoso: Una Economía del Conocimiento Circular

Las columnas que definen la nueva arquitectura del sistema de comunicación científica distan mucho de ser respuestas aisladas. Al contrario, forman un sistema integrado y sinérgico o auto-reforzante: el capital (de la QF) se transforma en propiedad intelectual (IP-NFTs), cuya creación y validación generan reputación (SBTs), y esta reputación, a su vez, confiere poder de gobierno (en las ARCs) para dirigir la asignación de futuro capital.

Este ciclo virtuoso crea una economía del conocimiento circular que logra algo fundamental: desagrega el monolítico "artículo de revista" en sus componentes de valor constituyentes. El sistema actual agrupa la idea, los datos, el código, la revisión y el sello de prestigio en un solo producto controlado por un intermediario. El nuevo sistema valora y recompensa cada uno de estos elementos de forma independiente (a través de SBTs e IP-NFTs), creando un mercado mucho más granular, eficiente y justo para las contribuciones científicas.

De esta forma, la verdadera promesa del nuevo sistema radica en su capacidad para ofrecer soluciones directas y estructurales a los problemas expuestos. No se trata de parches, sino de un rediseño fundamental de los incentivos y las infraestructuras, entre las que cabe destacar las siguientes:

- Respuesta a la Revisión por Pares No Remunerada: Las nuevas plataformas pueden crear sistemas de revisión por pares transparentes e incentivados. Mediante contratos inteligentes, los revisores pueden ser recompensados con tokens por su trabajo, y la calidad de sus revisiones, al ser pública, contribuye a su reputación en la cadena de bloques (*on-chain*). Esto transforma la revisión de una carga no reconocida en una contribución valorada y verificable.
- Respuesta a las Reglas de Sumisión Única y Mordaza: La infraestructura tecnológica descentralizada, como los servidores de preprints descentralizados (por ejemplo, utilizando *IPFS* o *Arweave*), permite la publicación instantánea, permanente y resistente a la censura de los resultados de la investigación. Esto anula el poder de las editoriales para actuar como guardianes (*gatekeepers*) y elimina los embargos que retrasan artificialmente la comunicación científica.

---

[167] https://www.wisdomtreeprime.com/blog/token-trust-how-soulbound-nfts-unlock-the-future-of-onchain-finance/

[168] https://www.cryptoaltruists.com/blog/5-desci-projects-revolutionizing-scientific-research-and-development

- Respuesta al Oligopolio y los Costos Exorbitantes: Al construir una infraestructura de publicación alternativa y de propiedad comunitaria, la nueva arquitectura socava el modelo de negocio del oligopolio editorial. Al eliminar intermediarios, los costos se reducen drásticamente, y el valor generado se retiene dentro de la comunidad científica en lugar de ser extraído como beneficio para los accionistas.
- Respuesta a la Financiación Opaca y la Crisis de Reproducibilidad: Las DAOs científicas, como *VitaDAO* (centrada en la longevidad) o *AthenaDAO* (centrada en la salud de la mujer), permiten que las decisiones de financiación sean tomadas por una comunidad global de expertos y partes interesadas, de forma transparente en la blockchain. Mecanismos como la QF amplifican las contribuciones de un gran número de pequeños donantes sobre las de unos pocos grandes, promoviendo proyectos con un amplio respaldo comunitario. La financiación retroactiva de bienes públicos permite recompensar proyectos que ya han demostrado su impacto, incentivando la creación de valor real en lugar de promesas. Además, la naturaleza inmutable de la blockchain asegura que los datos y métodos de investigación no puedan ser alterados, proporcionando una base sólida para la reproducibilidad.

A los efectos ilustrativos, a modo de resumen, la Tabla 2 muestra la arquitectura del Ecosistema Académico Descentralizado.

**Tabla 2: Plan Arquitectónico del Ecosistema Académico Descentralizado**

| Problema en el Sistema Actual (Parte I) | Principio Rector (Parte II) | Solución DeSci Propuesta (Parte III) | Mecanismo/Función Clave |
|---|---|---|---|
| **Revisión por pares opaca y sesgada por juntas controladas por editoriales.** | Gobernanza Democrática (Humanismo Digital) | DAOs de Revisión por Pares | Grupos de revisores con acceso basado en tokens o reputación, procesos de revisión transparentes, contribuciones incentivadas. |
| **Financiación de subvenciones centralizada y burocrática que ahoga la innovación.** | Expandir Capacidades (Econ. Humanista) | Modelo de Financiación Híbrido (QF + Misión) | Financiación Cuadrática para investigación de bienes públicos; DAOs estilo ARPA-H para proyectos ambiciosos ("moonshots"). |
| **Propiedad de la PI por parte de las editoriales, muros de pago.** | Orden Competitivo (Ordoliberalismo) | IP-NFTs con Mandatos de Acceso Abierto | Contratos inteligentes que automatizan el reparto de regalías y hacen cumplir las licencias abiertas. |
| **Métricas manipulables (Factor de Impacto, índice h).** | Valor Holístico (Econ. Humanista) | Sistema de Reputación con Tokens Vinculados al Alma (SBT) | Tokens no transferibles que registran diversas contribuciones (revisiones, datos, mentoría, etc.). |

# IV. Pruebas de Estrés: Forjando la Resiliencia y la Antifragilidad de la Nueva Comunicación Científica

El diseño de una arquitectura teóricamente superior es solo el primer paso. Un plan maestro que ignore la sociología de su adopción y los riesgos inherentes a sus propias herramientas está destinado al fracaso.
Esta cuarta parte somete la arquitectura propuesta a un riguroso análisis de vulnerabilidades, no para socavarla, sino para fortalecerla, identificando proactivamente los modos de fallo y diseñando mecanismos de mitigación a nivel de protocolo, en línea con el principio ordoliberal de una "policía de mercado" incorporada.

## IV.1 Prueba de Estrés Sociológica: La Inercia Institucional y el Problema del "Arranque"

La mayor barrera para la adopción de este nuevo sistema no es tecnológica, sino sociológica. Las universidades son instituciones con culturas y normas profundamente arraigadas que las hacen inherentemente resistentes al cambio [169], [170]. La resistencia del profesorado a adoptar nuevos modelos de publicación se basa en preocupaciones legítimas dentro del sistema de incentivos actual: el temor a que no confieran el mismo prestigio que las revistas tradicionales, afectando negativamente a sus carreras y evaluaciones [171], [172]. Esta renuencia está vinculada no sólo a aspectos económicos, sino a identidades profesionales, concepciones sobre la integridad pedagógica y la autonomía académica profundamente enraizadas en la tradición disciplinaria.
Esto crea el problema fundamental del "arranque del prestigio": ¿cómo puede la reputación *on-chain* (SBTs) adquirir valor si las instituciones *off-chain* (comités de contratación y promoción) que pretende suplantar no la reconocen? Este es un dilema de coordinación económica clásico que requiere romper la circularidad del equilibrio institucional existente. Los sistemas de ranking y acreditación académicos funcionan como mecanismos de señalización cuya legitimidad se sustenta en siglos de institucionalización; su reemplazo no puede ocurrir mediante sustitución unilateral, sino mediante procesos de "traducción institucional" deliberadamente orquestados.
Para superar esta barrera, es crucial diseñar "capas de traducción" y "puentes institucionales" [173]. Esto implica:
1. Formar coaliciones de universidades piloto que se comprometan formalmente a reconocer los SBTs en sus procesos de evaluación, promoción y contratación [174], [175]. Este modelo de adopción coordinada reduce el riesgo individual para cada institución y genera economías de escala en la infraestructura de validación. La evidencia de adopción institucional temprana en tecnologías educativas demuestra que la colaboración entre universidades es más efectiva que la adopción aislada.

---

[169] https://www.iied.org/sites/default/files/pdfs/migrate/10763IIED.pdf

[170] https://www.researchgate.net/publication/236771127_The_Effect_of_Institutional_Culture_on_Change_Strategies_in_Higher_Education

[171] https://pmc.ncbi.nlm.nih.gov/articles/PMC12365406/

[172] https://twu-ir.tdl.org/bitstreams/f08c0992-5ee2-407d-bf32-7ca7f6e42fdc/download

[173] https://www.researchgate.net/publication/281505256

[174] https://doi.org/10.3390/app9122400

[175] https://doi.org/10.3389/fbloc.2025.1641294

2. Desarrollar herramientas de software que agreguen y presenten la reputación *on-chain* en formatos legibles y convincentes para los comités tradicionales (un "CV aumentado") [176]. Este requisito técnico está íntimamente relacionado con el problema de traducción: los SBTs deben convertirse en artefactos inteligibles dentro del lenguaje institucional establecido, preservando su valor verificable mientras se mapean a criterios de evaluación reconocidos. La investigación en credenciales verificables blockchain sugiere que la interoperabilidad entre sistemas de identidad descentralizados y registros académicos tradicionales es técnicamente viable, pero requiere estándares consensuados.
3. Lanzar programas de financiación que den prioridad explícita a los investigadores que operan y contribuyen dentro del nuevo sistema, creando un incentivo tangible para la adopción temprana [177]. Este mecanismo invierte deliberadamente en actores con suficiente flexibilidad institucional (jóvenes investigadores, equipos interdisciplinarios, grupos de investigación de vanguardia) para pilotar nuevos modelos antes de que se generalicen. La economía de redes sugiere que estos subsidios a la adopción temprana son críticos para romper equilibrios de coordinación negativos.

## IV.2 Prueba de Estrés Económica: La Hiper-Financiarización y la Especulación

La introducción de la tokenómica y los IP-NFTs crea el riesgo de una hiper-financiarización de la investigación, donde la ciencia se trata como una clase de activo especulativo en lugar de un bien público [178], [179]. Los mercados de IP-NFTs podrían convertirse en burbujas volátiles, desconectadas del valor científico subyacente, replicando los peores excesos de las finanzas descentralizadas (DeFi) [180], [181], [182]. La historia económica documenta extensamente cómo los mecanismos de tokenización conducen sistemáticamente a ciclos especulativos cuando faltan amortiguadores institucionales y regulatorios; el colapso del mercado de NFTs entre 2021 y 2023 ejemplifica este patrón, con volúmenes de negociación cayendo un 97% en nueve meses.

Este riesgo es particularmente agudo en contextos de investigación, donde el verdadero valor de un descubrimiento sólo se establece después de ciclos largos de validación y reproducibilidad. La brecha temporal entre la tokenización y la validación científica crea una ventana para el arbitraje especulativo; los primeros compradores de tokens de PI podrían beneficiarse enormemente si los precios se desvinculan de la calidad científica subyacente, creando incentivos para la emisión acelerada de tokens de bajo valor.

La mitigación de este riesgo requiere una *Ordnungspolitik* a nivel de protocolo, incorporando reglas en los contratos inteligentes para desalentar la especulación [183]. Como se mencionó en la parte II, la escuela ordoliberal propone que las órdenes económicas requieren marcos legales superiores que estructuren las interacciones de mercado antes de que estas se materialicen. En el contexto de las cadenas de bloques, esto significa codificar restricciones desincentivadoras de la especulación directamente en los mecanismos de consenso y los contratos inteligentes.

---

[176] https://www.researchgate.net/publication/363063912

[177] https://www.ulam.io/blog/how-decentralized-science-is-revolutionizing-research

[178] https://d-nb.info/1024714179/34

[179] https://pmc.ncbi.nlm.nih.gov/articles/PMC4719771/

[180] https://medium.com/paradigm-research/intellectual-property-in-science-the-potential-advantages-of-nfts-907167b6af08

[181] https://committees.parliament.uk/publications/41611/documents/205745/default/

[182] https://arxiv.org/html/2312.01018v1

[183] https://mitsloan.mit.edu/ideas-made-to-matter/decentralized-finance-4-challenges-to-consider

Esto implica:
1. Períodos de vesting y lock-up para los tokens de los inversores iniciales y equipos de desarrollo, asegurando un compromiso a largo plazo [184], [185]. Estos mecanismos obligan a los tenedores tempranos a mantener posiciones durante períodos definidos (típicamente 6-24 meses), eliminando el incentivo de "pump-and-dump" y alineando los intereses con el desempeño a largo plazo del protocolo. Proyectos como *Compound* e *Uniswap* han implementado lockups con éxito como defensa contra manipulación de gobernanza mediante "flash loan attacks".
2. Modelos de staking de reputación, donde para participar en la financiación de un proyecto, los usuarios deben bloquear no solo capital (tokens) sino también SBTs relevantes, vinculando la inversión financiera al mérito y la experiencia demostrados [186]. Este mecanismo hace que el costo de oportunidad de la especulación sea proporcional a la credibilidad académica en juego; un inversor especulador tendría que sacrificar su reputación científica para operar, creando un freno para la conducta delictiva.
3. Curvas de vinculación (bonding curves) para la tokenización de la PI que estabilicen los precios y recompensen a los tenedores a largo plazo en lugar de a los especuladores a corto plazo [187], [188], [189]. Las curvas de vinculación son funciones matemáticas que definen la relación entre el precio de un token y su suministro en circulación; permiten la determinación automática de precios basada en demanda agregada. Un diseño de curva exponencial favorecería a los compradores tempranos, pero también penalizaría a los vendedores rápidos a través de mecanismos de depreciación temporal, creando así un desincentivo natural para el ciclo especulativo. La investigación en "*risk-adjusted bonding curves*" demuestra que estos mecanismos pueden integrar predicciones sobre la calidad del proyecto e incorporarlas dinámicamente en la fijación de precios.

## IV.3 Prueba de Estrés Ética: El Espectro de la Vigilancia y el "Crédito Social"

Los Tokens Soulbound, aunque potentes, presentan graves riesgos para la privacidad. Un registro público e inmutable de cada contribución científica podría crear un sistema de vigilancia distópico, una especie de "sistema de crédito social" para académicos donde cada acción es permanentemente registrada y juzgada [190], [191]. La irreversibilidad de los SBTs también plantea problemas sobre cómo corregir errores o revocar credenciales emitidas injustamente [192].

El sistema de crédito social chino, aunque implementado en un contexto autoritario, ilustra los riesgos inherentes a los registros inmutables de comportamiento individual: la descentralización de decisiones de punición, la imposibilidad de borrado de datos y la permanencia de juicios negativos crean un régimen de control que niega a los individuos la oportunidad de redención o corrección de

---

errores administrativos. Aunque los SBTs en contextos académicos occidentales operarían en marcos democráticos distintos, la arquitectura tecnológica subyacente replicaría estas vulnerabilidades si no se diseña cuidadosamente con salvaguardas.

El diseño del sistema debe incorporar proactivamente tecnologías que preserven la privacidad, tales como las siguientes:

1. Implementación obligatoria de Pruebas de Conocimiento Cero (ZKPs), permitiendo a un investigador demostrar hechos sobre su reputación (por ejemplo: "*he realizado 23 revisiones para revistas de economía*") sin revelar los datos subyacentes [193], [194], [195]. Las pruebas de conocimiento cero son protocolos criptográficos que permiten a una parte demostrar que una afirmación es verdadera sin revelar información adicional más allá de la validez de la afirmación. En contextos académicos, un investigador podría demostrar que posee suficiente reputación en revisión de pares sin exponer el contenido específico de sus reseñas ni la identidad de los autores evaluados. La investigación en zk-SNARKs (Zero-Knowledge Succinct Non-Interactive Arguments of Knowledge) demuestra que estos protocolos pueden aplicarse a sistemas de credenciales descentralizadas con sobrecarga computacional mínima.

2. Estándares de datos que separen la prueba criptográfica *on-chain* de los metadatos sensibles, que se almacenarían fuera de la cadena en redes de almacenamiento descentralizado o bóvedas de datos privadas [196], [197]. Este diseño híbrido preserva los beneficios de inmutabilidad y verificación de blockchain para el núcleo criptográfico de la credencial, mientras que los detalles sensibles (comentarios de revisión, datos biométricos, información de salud) permanecen bajo control del individuo. Las redes de almacenamiento descentralizado como IPFS pueden ser combinadas con encriptación local para garantizar que ni siquiera los operadores de la red pueden acceder a datos sensibles.

3. Sistemas de revocación y apelación gobernados democráticamente por las ARCs para corregir SBTs emitidos injustamente, asegurando la responsabilidad humana en el sistema [198]. La revocación de credenciales en sistemas blockchain presenta desafíos técnicos únicos, ya que la inmutabilidad del registro impide simplemente "borrar" datos incorrectos; las soluciones propuestas incluyen mecanismos de revocación criptográfica (donde un SBT puede ser marcado como revocado mediante una transacción que no modifica el registro original) y estructuras de gobernanza que permitan anulaciones mediante votación comunitaria con apelaciones judiciales. La investigación en revocación de certificados blockchain demuestra que estos sistemas pueden implementarse con latencia de minutos en lugar de días, permitiendo respuestas rápidas a denuncias de falsificación o injusticia.

## IV.4 Prueba de Estrés Política: La Ilusión de la Descentralización

Existe un riesgo real de que la "descentralización" sea una simple fachada, ocultando nuevas formas de centralización en manos de un pequeño grupo de desarrolladores clave, grandes poseedores de tokens ("ballenas") o capitalistas de riesgo, simplemente reemplazando a los antiguos guardianes por otros nuevos y menos visibles [199], [200]. Este fenómeno es particularmente visible en protocolos de las cadenas de bloques que comenzaron con aspiraciones radicales, pero evolucionaron hacia estructuras oligárquicas de facto: *Bitcoin* permite de facto la dominación de la toma de decisiones por desarrolladores principales; *Ethereum* ha concentrado poder de voto significativamente en tenedores institucionales de *large stakes*; numerosas DAOs han experimentado captura por "ballenas" que poseen suficientes tokens para bloquear cambios de protocolo que amenacen sus intereses.

---

[193] https://www.webopedia.com/crypto/learn/what-is-a-soulbound-token/

[194] https://store.aicerts.ai/blog/the-rise-of-soulbound-tokens-sbts-a-new-frontier-for-non-transferable-nfts/

[195] https://www.leewayhertz.com/soulbound-tokens/

[196] https://www.itm-conferences.org/articles/itmconf/pdf/2023/06/itmconf_icdsac2023_06002.pdf

[197] https://www.wisdomtreeprime.com/blog/token-trust-how-soulbound-nfts-unlock-the-future-of-onchain-finance

[198] https://www.sciencedirect.com/science/article/pii/S016740482100033X

Al objeto de garantizar una descentralización genuina, el sistema debería diseñarse con mecanismos de gobernanza progresiva y salidas creíbles, a saber:

1. Descentralización progresiva, un camino claro y programado desde un equipo central inicial hacia una gobernanza totalmente comunitaria a medida que el protocolo madura [201], [202].

   Este *roadmap* debe ser codificado en contratos inteligentes con hitos explícitos: por ejemplo, en el mes 12 de operación, el poder de veto centralizado se reduce de 51% a 40%; en el mes 24, a 20%; en el mes 36, se elimina completamente. La investigación en gobernanza blockchain demuestra que las transiciones abruptas son propensas a captura; las transiciones graduales permiten que las comunidades se auto-organicen y desarrollen estructuras descentralizadas robustas antes de que sea crítico.

2. Constituciones de DAO robustas que incluyan fuertes protecciones para los derechos de las minorías y mecanismos de seguridad como bloqueos de tiempo (time-locks) y carteras multifirma (multi-signature wallets) para prevenir ataques a la gobernanza [203], [204], [205].

   Un timelock es un contrato inteligente que introduce un retraso obligatorio entre la aprobación de una propuesta de cambio de protocolo y su ejecución, permitiendo que la comunidad detecte y rechace propuestas maliciosas antes de que tomen efecto. Los timelock efectivos requieren períodos de retraso que equilibren la capacidad de respuesta rápida con la detección de amenazas (típicamente 2-7 días); períodos más cortos facilitan ataques, períodos más largos ralentizan la gobernanza legítima.

   Las carteras multifirma requieren múltiples claves privadas para autorizar transacciones críticas (como transferencias de tesorería o cambios de parámetros de protocolo), distribuyendo el poder de veto entre múltiples partes; en una configuración 3-de-5, por ejemplo, ningún actor individual puede imponer su voluntad sin al menos dos aliados.

   Las constituciones de DAO también deben establecer "vetos de minoría": mecanismos que permitan que grupos relativamente pequeños de participantes (típicamente 10-30% de la población votante) bloqueen cambios que consideren que violarían principios fundamentales del protocolo. Aunque esto reduce la velocidad de cambio, también previene que mayorías capturadas por ballenas o coaliciones maliciosas reescriban unilateralmente las reglas del juego.

3. Interoperabilidad y estándares abiertos por diseño, lo que garantiza que, si un protocolo es capturado por intereses centralizadores, la comunidad puede "bifurcarlo" (fork) y migrar a una alternativa. Esta amenaza de salida creíble es el máximo garante contra la tiranía del protocolo [206], [207].

---

[199] https://wfm-igp.org/federalist-paper/the-dao-and-the-dao-finding-a-path-to-govern-the-world/

[200] https://www.researchgate.net/publication/390230282_DAO_as_digital_governance_tool_for_collaborative_housing

[201] https://www.bruegel.org/policy-brief/decentralised-finance-good-technology-bad-finance

[202] https://pmc.ncbi.nlm.nih.gov/articles/PMC11839390/

[203] https://pdfs.semanticscholar.org/90b7/15129b5cc2ef3b7929c515519379b832a5d0.pdf

[204] https://arxiv.org/html/2312.01018v1

[205] https://www.researchgate.net/publication/375240188_Decentralised_Finance_DeFi_a_critical_review_of_related_risks_and_regulation

Un fork es una división en la historia de transacciones de un blockchain: cuando la comunidad no está de acuerdo sobre cambios propuestos, puede crear una versión alternativa del protocolo que prosigue con un conjunto distinto de reglas; el fork más famoso es Bitcoin Cash (BCH), que divergió de Bitcoin (BTC) en 2017 sobre desacuerdos sobre tamaño de bloque y gobernanza.

La facilidad de *forking* depende de tres factores técnicos: (a) transparencia de especificación del protocolo, asegurando que cualquier desarrollador puede construir una implementación alternativa sin acceso privilegiado a código propietario; (b) interoperabilidad de datos, permitiendo que DAOs migren su estado (registros de miembros, tesorería, histórico de votaciones) entre versiones del protocolo sin perder datos; (c) portabilidad de identidad, permitiendo que los usuarios transfieran sus SBTs, reputación y credenciales verificables entre versiones del protocolo. Estos requisitos están implementados técnicamente a través de estándares abiertos como ERC (*Ethereum Request for Comments*) para tokens y protocolos de puente descentralizados que permiten que activos se trasladen entre cadenas incompatibles.

La teoría económica de la descentralización sugiere que el poder de salida creíble (la amenaza de que los usuarios pueden abandonar) es estructuralmente más fuerte que el poder de voz (la capacidad de votar sobre cambios internos) como restricción sobre el comportamiento oportunista de los gobernantes. Cuando un fork es técnicamente factible y socialmente legítimo, incluso una minoría significativa puede disciplinar a la mayoría mediante la amenaza de fraccionamiento.

# V. Hoja de Ruta para la Transición: De la Arquitectura a la Realidad Institucional

El diseño de un nuevo orden, por muy coherente que sea, sigue siendo un ejercicio teórico si no se traza un camino plausible desde el sistema actual hasta el futuro imaginado. La transición de un paradigma institucional arraigado a uno emergente es el desafío más formidable, plagado de problemas de acción colectiva, inercia y riesgos para los pioneros. Esta sección detalla una hoja de ruta estratégica y multifásica, diseñada para navegar esta complejidad, arrancar la legitimidad y escalar la adopción de la Nueva Comunicación Científica de manera deliberada y resiliente.

## V.1 Fase I: Génesis y Arranque Institucional (Años 1-3) - La Estrategia del Nicho Protegido

El objetivo principal de esta fase inicial es resolver el "Dilema del Pionero" creando un entorno de bajo riesgo o "nicho protegido" donde las nuevas reglas del juego puedan ser adoptadas y validadas por un grupo inicial de actores comprometidos. Para ello cabe llevar a cabo, entre otras, las acciones siguientes:
- Acción 1: Formación de la Alianza Fundacional. El primer paso es la creación de un consorcio de vanguardia compuesto por un número selecto (por ejemplo, 5-10) de universidades con visión de futuro, agencias de financiación progresistas y sociedades científicas influyentes. Inspirado en modelos de acción colectiva exitosos como SCOAP en física de partículas y las coaliciones detrás de la Declaración de San Francisco sobre la Evaluación de la Investigación (DORA), este consorcio firmará un pacto fundacional. Este pacto comprometerá formalmente a las instituciones miembros a reconocer las contribuciones dentro del nuevo ecosistema (como los SBTs) como criterios válidos y valiosos en sus procesos internos de contratación, promoción y financiación.
- Acción 2: Desarrollo del Protocolo Mínimo Viable (MVP) y la Constitución Inicial. Paralelamente, un equipo de desarrollo del núcleo, financiado por la Alianza, construirá la infraestructura técnica esencial. El MVP no necesita replicar todo el sistema, sino centrarse en los componentes clave para arrancar el ciclo de reputación: un sistema de identidad descentralizada ("Alma"), mecanismos para la emisión de SBTs para un conjunto limitado de acciones (por ejemplo, revisión por pares abierta, publicación de preprints con datos abiertos) y una estructura básica de ARC/DAO para la gobernanza del piloto. Crucialmente, la constitución inicial de esta DAO debe codificar el principio de descentralización progresiva, incluyendo una cláusula de extinción (*sunset clause*) que retire irrevocablemente los poderes administrativos especiales del equipo fundador una vez que se alcancen hitos de madurez predefinidos, garantizando un compromiso creíble con la soberanía comunitaria a largo plazo.
- Acción 3: Lanzamiento de Programas de Financiación Piloto Catalizadores. Las agencias de financiación de la Alianza lanzarán convocatorias de subvenciones diseñadas explícitamente para incentivar la adopción. Estas convocatorias podrían requerir que las solicitudes incluyan un portafolio de SBTs, que los resultados se publiquen como IP-NFTs de acceso abierto, o que se experimente con la Financiación Cuadrática para asignar una parte de los fondos. Al vincular directamente la financiación con la participación en el nuevo sistema, estas agencias alteran drásticamente el cálculo de riesgo-recompensa para los investigadores pioneros.
- Acción 4: Creación del Oráculo de Reputación v1.0. Para tender un puente entre el mundo *on-chain* y el *off-chain*, la Alianza establecerá la primera versión del "Oráculo de Reputación". Este organismo, gobernado por representantes de las instituciones miembros, tendrá la tarea de interpretar los portafolios de SBTs y generar informes cualitativos y narrativos que sean legibles y significativos para los comités de evaluación tradicionales. Su legitimidad no derivará de la tecnología, sino del prestigio de las instituciones que lo respaldan.

---

[206] https://papers.ssrn.com/sol3/papers.cfm?abstract_id=2709713

[207] https://papers.ssrn.com/sol3/papers.cfm?abstract_id=2852691

## V.2 Fase II: Expansión y Legitimación (Años 3-7) - Cruzando el Abismo

Con un nicho protegido y funcional, el objetivo de la segunda fase es escalar la adopción más allá de los primeros innovadores para alcanzar una masa crítica y lograr efectos de red. Para lograrlo se precisaría, entre otras, las acciones siguientes:
- Acción 1: Escalado de la Red Institucional y Estandarización. La Alianza Fundacional se expandirá activamente, utilizando los casos de éxito y los datos del piloto para reclutar a una segunda cohorte de instituciones "seguidoras rápidas". Se desarrollarán y promoverán estándares abiertos para la interoperabilidad de los SBTs y los IP-NFTs, asegurando que el ecosistema no se fragmente en silos incompatibles.
- Acción 2: Desarrollo de Mercados de Conocimiento y Herramientas Avanzadas. La infraestructura se expandirá para incluir funcionalidades más sofisticadas. Se lanzarán las primeras plataformas de mercado para IP-NFTs, permitiendo la financiación y comercialización de la investigación traslacional. Se implementarán mecanismos de Financiación Cuadrática a mayor escala. Se desarrollarán herramientas de software que se integren con los sistemas universitarios existentes (como los sistemas CRIS y de recursos humanos) para reducir la fricción en la adopción institucional.
- Acción 3: Institucionalización de la Gobernanza Bicameral y Federalista. A medida que las ARCs crezcan en tamaño y diversidad, su gobernanza evolucionará. Se implementarán modelos bicamerales, separando una "Cámara Plural" (basada en Votación Cuadrática para decisiones de base amplia) de una "Cámara Epistémica" (compuesta por miembros de alta reputación para salvaguardar los principios a largo plazo). Se fomentará un modelo federalista, donde las ARCs de subdisciplinas puedan operar con autonomía dentro de ARCs más amplias, permitiendo la experimentación con reglas locales sin comprometer la estabilidad del sistema global.
- Acción 4: Fomento de la Interoperabilidad y la "Forkeabilidad". Para consagrar el principio ordoliberal de prevenir la concentración de poder, el desarrollo del protocolo priorizará la interoperabilidad y la "forkeabilidad" (la capacidad de la comunidad para bifurcar el código y crear una versión competidora): la amenaza creíble de una bifurcación actúa como el máximo control sobre cualquier intento de captura del protocolo, asegurando que la gobernanza permanezca responsable ante la comunidad.

## V.3 Fase III: Consolidación y Soberanía Plena (Año 8 y más allá) - El Nuevo Statu Quo

El objetivo de la fase final es la consolidación del nuevo ecosistema como la infraestructura por defecto para la ciencia, relegando al sistema heredado a un papel secundario. Entre las acciones a desarrollar para su logo cabe destacar las siguientes:
- Acción 1: Transferencia Total de la Soberanía del Protocolo. Los hitos definidos en la constitución inicial se alcanzan, activando la cláusula de extinción. El control total sobre la meta-gobernanza del protocolo (la capacidad de enmendar sus reglas fundamentales) se transfiere irrevocablemente a las estructuras de gobernanza comunitarias (las ARCs federadas). El equipo fundador y la Alianza original se convierten en participantes en igualdad de condiciones con el resto de la comunidad.
- Acción 2: El Sistema Heredado como "Capa de Compatibilidad". En este punto, los efectos de red del nuevo sistema son tan fuertes que las instituciones y editoriales heredadas que no han hecho la transición se enfrentan a la irrelevancia. Su estrategia de supervivencia será integrarse en la Nueva Mancomunidad. Las revistas tradicionales podrían reconvertirse en ARCs de curación especializadas, emitiendo SBTs por su sello editorial, pero operando dentro de las reglas y la economía del nuevo ecosistema. El Oráculo de Reputación ya no traduce lo nuevo para lo viejo, sino que ayuda a las contribuciones del sistema heredado a ser representadas en el nuevo.
- Acción 3: La Mancomunidad como Infraestructura Pública Global. El sistema alcanza la madurez como un bien público digital global, análogo a los protocolos centrales de Internet. Su mantenimiento y desarrollo continuo se financian a través de una combinación de ingresos del protocolo (por ejemplo, una pequeña tasa sobre las transacciones de IP-NFTs), dotaciones filantrópicas y financiación pública de consorcios gubernamentales, similar a cómo se financian infraestructuras como *CERN* o el *Protocol Guild de Ethereum*.
- Acción 4: Cultivo de una Cultura de Evolución Constitucional Continua. La antifragilidad del sistema a largo plazo no reside en la perfección de su diseño inicial, sino en su capacidad para evolucionar. La actividad principal de la gobernanza comunitaria se desplaza hacia la meta-gobernanza: el debate, la propuesta y la ratificación de enmiendas a la "constitución" del protocolo para adaptarse a nuevos desafíos tecnológicos, sociales y epistémicos. El sistema se convierte en un experimento inacabado, una constitución viviente para la ciencia.

# VI. Manifiesto Constituyente: Hacia una Nueva República de las Ideas

## VI.1 El Cierre del Arco: De la Diagnosis a la Fundación

Este documento comenzó con una premisa audaz: el orden académico actual no está simplemente fallando; está fundamentalmente quebrado, y el momento exige la fundación de una nueva "República de las Ideas". Las partes precedentes han cumplido con el deber de esta ambición. Hemos diagnosticado la enfermedad del viejo orden (Parte I), establecido los principios filosóficos de uno nuevo (Parte II), diseñado su arquitectura técnica (Parte III), sometido ese diseño a pruebas de estrés (Parte IV) y trazado un camino plausible para su construcción (Parte V).
Esta sección final, por tanto, no es un resumen. Es un acto de fundación. Cierra el arco narrativo no repitiendo lo dicho, sino transformando el análisis en un manifiesto constituyente. Lo que sigue son los principios y la llamada a la acción.

## VI.2 El Momento Constitucional de la Ciencia

Nos encontramos en una encrucijada histórica, un *"momento constitucional"* para la ciencia. El antiguo pacto social que gobernaba la producción de conocimiento está irrevocablemente roto. Sus instituciones, nacidas en la era de la imprenta y la escasez, se han vuelto en contra de su propósito original. En lugar de acelerar el descubrimiento, lo ralentizan. En lugar de recompensar el rigor, incentivan la superficialidad. En lugar de servir al bien público, han sido capturadas por intereses privados que extraen rentas del intelecto humano colectivo. La crisis de replicación no es un fallo técnico; es la bancarrota moral de un orden agotado.
Continuar con reformas incrementales es abdicar de nuestra responsabilidad. Es aplicar parches a un edificio cuyos cimientos se han podrido. La confluencia de un fracaso sistémico tan profundo con la aparición de un nuevo paradigma tecnológico —la descentralización— nos otorga una oportunidad generacional única: no la de reformar, sino la de refundar.
La tarea que tenemos por delante no es la de litigar los términos de nuestra servidumbre a un orden fallido, sino la de declarar nuestra independencia y redactar la constitución de uno nuevo.

## VI.3 Los Principios Fundamentales de la República

Toda república duradera se funda sobre principios inalienables. La nuestra, la República de las Ideas, se erigirá sobre la trinidad filosófica que ha guiado este análisis:
- Artículo I (Del Orden Justo): Se establece un gobierno de protocolos, no de oligarcas. Inspirados por el Ordoliberalismo, construiremos una *Ordnungspolitik* para la ciencia, una constitución económica codificada que garantice la competencia leal, prevenga la concentración de poder y asegure que las reglas del juego sean transparentes y estén al servicio de todos sus ciudadanos, no de una élite editorial o una tecno-aristocracia.[14]
- Artículo II (Del Propósito Humano): El *telos*, el fin último de esta República, será el florecimiento humano. Guiados por la Economía Humanista de Amartya Sen, mediremos nuestro éxito no por la acumulación de métricas vacías, sino por la expansión de las capacidades humanas: la libertad real de los científicos para investigar las preguntas que importan, la capacidad de la sociedad para acceder al conocimiento que necesita y el fomento de una prosperidad compartida basada en el descubrimiento.[21]

- Artículo III (De la Soberanía del Ciudadano): La tecnología estará al servicio del científico, y no al revés. Siguiendo los mandatos del Humanismo Digital, diseñaremos nuestras herramientas para potenciar la autonomía, la dignidad y la creatividad humanas. Cada ciudadano de esta república tendrá soberanía sobre su identidad, su reputación y sus contribuciones, protegido por una declaración de derechos digitales inmutables.

## VI.4 La Llamada a los Fundadores: Un Manifiesto para la Acción

Esta constitución no se escribirá sola. Exige una generación de fundadores. Esta es una llamada a los ciudadanos de la República aún no nacida:
- A los Investigadores Pioneros: El "Dilema del Pionero" no es una barrera, sino la definición misma del coraje revolucionario. Los primeros en adoptar estos nuevos protocolos, en construir su reputación *on-chain* y en desafiar la tiranía del Factor de Impacto, no serán meros "early adopters"; serán los firmantes de una nueva declaración de independencia intelectual. Vuestro riesgo es el precio de la libertad para las generaciones venideras.
- A las Instituciones de Vanguardia: Las universidades, agencias de financiación y sociedades científicas que se unan a la "Alianza Fundacional" no estarán participando en un proyecto piloto. Estarán convocando la Convención Constitucional de la ciencia del siglo XXI. Vuestra tarea es crear el "nicho protegido" que servirá de refugio seguro para los primeros revolucionarios y de laboratorio para las nuevas leyes de la República.
- A los Arquitectos del Protocolo: Vuestra vocación no es la de meros ingenieros, sino la de legisladores constitucionales. Cada línea de código que escribís, cada mecanismo de gobernanza que diseñáis es un artículo de esta nueva constitución. Abrazad la descentralización progresiva no como una opción técnica, sino como un imperativo moral: el compromiso de ceder el poder, de programar vuestra propia obsolescencia para que la soberanía resida, en última instancia, en la comunidad.[59]

La elección que enfrentamos no es entre sistemas de publicación, sino entre épocas. Es la elección entre perpetuar una era de escasez artificial, extracción de rentas y parálisis institucional, o inaugurar una era de abundancia, creación de valor y descubrimiento acelerado. La Hoja de Ruta aquí trazada no es un conjunto de recomendaciones pasivas; es un plan de campaña.

## VI.5 Advenimiento de una Constitución Viviente para la Ciencia

La República de las Ideas no será una utopía estática. Su antifragilidad no residirá en la perfección de su código fundacional, sino en su capacidad para la autocrítica y la evolución. Será, por diseño, un experimento inacabado, una constitución viviente cuya legitimidad deberá ser renovada por cada generación de ciudadanos. Los mecanismos de meta-gobernanza, las pruebas de estrés y los contrapesos institucionales no son meras salvaguardias; son la encarnación de la humildad socrática en el corazón de la empresa científica: el reconocimiento de que nuestro conocimiento siempre es provisional y de que nuestras instituciones deben ser tan falsificables como nuestras teorías.

La tarea es monumental, pero la alternativa —la continua decadencia del orden actual— es impensable. Estamos interpelados a ser los arquitectos de una nueva infraestructura para la verdad, un procomún del conocimiento global que sea digno de la inteligencia y la creatividad de la especie humana.

La construcción de la Nueva República de las Ideas ha comenzado.

Vale.